\documentclass{aa}
\usepackage[dvips]{graphics}
\usepackage{longtable}
\setlongtables

\def\mic{\,\mu{\rm m}}

\begin{document}

\title{Explanatory Supplement of the ISOGAL-DENIS \\
Point Source Catalogue 
\thanks{This is paper no. 18 in a refereed journal based on data from
the ISOGAL project}\fnmsep
\thanks{Based on observations with ISO,
an ESA project with instruments funded by ESA Member States (especially
the PI countries: France, Germany, the Netherlands and the United Kingdom)
and with the participation of ISAS and NASA; and on DENIS observations
collected at the European Southern Observatory, Chile}}


\author{F. Schuller\inst{1} \and S. Ganesh\inst{2,1} \and M. Messineo\inst{3}
\and A. Moneti\inst{1} \and J.A.D.L. Blommaert\inst{4} \and C. Alard\inst{1,5}
\and B. Aracil\inst{1} \and M.-A. Miville-Desch\^enes\inst{6} \and
A. Omont\inst{1} \and M. Schultheis\inst{1} \and G. Simon\inst{5} \and
A. Soive\inst{1} \and L. Testi\inst{7}}
\offprints{F. Schuller, schuller@iap.fr}
\institute{
Institut d'Astrophysique de Paris, CNRS, 98 bis Bd Arago,
F-75014 Paris, France
\and
Physical Research Laboratory, Navarangpura,
Ahmedabad 380009, India
\and
Leiden Observatory, University of Leiden, P.O. Box 9513,
2300 RA Leiden, The Netherlands
\and
Instituut voor Sterrenkunde, K. U. Leuven, Celestijnenlaan 200 B,
B-3001 Leuven, Belgium
\and
GEPI, Observatoire de Paris, 61, av. de l'Observatoire,
F-75014 Paris, France
\and
Laboratoire de radioastronomie millim\'etrique, Ecole Normale
Sup\'erieure \& Observatoire de Paris, France
\and
Osservatorio Astrofisico di Arcetri, Largo E. Fermi, 5,
50125 Firenze, Italy
}

\date{Received xxxx / Accepted xxxx}

\abstract{We present version 1.0 of the ISOGAL--DENIS Point Source Catalogue
(PSC), containing more than 100,000 point sources detected at 7 and/or
15~$\mu$m in the ISOGAL survey of the inner Galaxy with the ISOCAM instrument
on board the {\em Infrared Space Observatory} (ISO). These sources are
cross-identified, wherever possible, with near-infrared (0.8---2.2~$\mu$m)
data from the {\em DENIS} survey. The overall surface covered by the ISOGAL
survey is about 16 square degrees, mostly (95$\%$) distributed near the
Galactic plane ($\vert$b$\vert \la 1^{\circ}$), where the source extraction
can become confusion limited and perturbed by the high background emission.
Therefore, special care has been taken aimed at limiting the photometric error
to $\sim 0.2\,$magnitude down to a sensitivity limit of typically $10\,mJy$.
The present paper gives a complete description of the entries and the
information which can be found in this catalogue, as well as a detailed
discussion of the data processing and the quality checks which have been
completed. The catalogue is available via the VizieR Service at the Centre
de Donn\'ees Astronomiques de Strasbourg
(CDS, http://vizier.u-strasbg.fr/viz-bin/VizieR/)
and also via the server at the Institut d'Astrophysique
de Paris (http://www-isogal.iap.fr/).
A more complete version of this paper, including a detailed
description of the data processing, is available in electronic
form through the ADS service.
}

\authorrunning{F. Schuller et al.}
\titlerunning{Explanatory Supplement of the ISOGAL-DENIS PSC}
\maketitle



{\it Note}: All subsections of Sect. 3 and 4 are more developed in the
electronic version of this paper.

\section{Introduction}

The ISOGAL survey is the
most sensitive mid-infrared wide-field survey dedicated to the inner Galaxy
(see the accompanying paper Omont et al. \cite{Omont2002} and
references therein for a review of its scientific goals and results).
The large amount of ISO observations collected, in combination with the
near-infrared data of the DENIS survey, has resulted in the production of a
catalogue of 10$^5$ point sources, the PSC. The first scientific results
obtained include studies of the Galactic structure, analysis of the 
stellar populations comprising completely
detected AGB stars with their mass-loss in particular fields (P\'erault et
al. \cite{perault}; Omont et al. \cite{omont99}; Glass et al. \cite{glass99};
Ojha et al. \cite{ref-ojha}), characterisation of interstellar extinction
(Jiang et al. \cite{ref-jiang}), of infrared dark clouds (Hennebelle
et al. \cite{hennebel}), and of young stellar objects (Felli
et al. \cite{felli00} and \cite{felli02}; Schuller \cite{phd-schuller}).

A total of $\sim$16 square degrees of the inner Galactic disk
($\vert$b$\vert \la 1^{\circ}$) were observed, with strong emphasis on
the inner Galactic bulge, at wavelengths of 7 and 15~$\mu$m, with a pixel
scale of usually 6'' and sometimes 3'', down to a sensitivity limit of
typically 10~mJy. A total of $\sim 250\,$hours of ISO
time were used, making ISOGAL one of the largest programs performed by ISO.
For the southern sky the results were combined with the $I$, $J$, $K_{\rm s}$
(effective wavelengths equal to 0.79, 1.22 and 2.14~$\mu$m) ground-based
data from the DENIS survey (Epchtein et al. \cite{DENIS}, \cite{ref-DENIS2})
in order to produce
an (up to) 5-wavelength catalogue of point sources. Given the emphasis of
ISOGAL on the inner Galactic regions, the DENIS coverage is available
for 95\% of the fields surveyed with ISOCAM.

As a comparison, the IRAS satellite, which made a breakthrough in the
infrared window in 1983,
performed an all sky survey resulting in a 2.5$\times$10$^5$ point source
catalogue, with a typical sensitivity (or 90\% completeness level) around
0.5~Jy in low source density regions and at the shortest wavelengths.
The four IRAS bands were centred at 12, 25, 60 and 100~$\mu$m, thus covering
the mid- to far-infrared range, with a spatial resolution ranging from
less than 1' at 12~$\mu$m to about 4' at 100~$\mu$m.
The sensitivity of ISOCAM is about two orders of magnitude better than that
provided by the IRAS detectors at 12~$\mu$m in the high source density
regions (thus in particular in the Galactic plane). Indeed, as explained in
the IRAS Explanatory Supplement (Section VIII), the typical 50\% completeness
limit flux density was about 1~Jy at 12 and 25~$\mu$m in the Galactic Plane,
and even brighter at longer wavelengths.

More recently, the MSX (Midcourse Space Experiment, see Mill et al.
\cite{msx-overview} for an overview) mission surveyed the complete
Galactic Disk in the range $\vert$b$\vert \leq 5^{\circ}$ in the
mid-infrared, using a 33~cm aperture telescope called SPIRIT III
(Price et al. \cite{ref-MSX}). Six bands between 4 and 25~$\mu$m were surveyed
simultaneously at a spatial resolution of $\sim$~18''. The most sensitive
band was the A band, centred at 8.3~$\mu$m, for which
the present point source sensitivity limit is about 0.1~Jy. The survey
of the Galactic Plane has presently resulted in a catalogue of
3.2$\times$10$^5$ sources (Price et al. \cite{ref-MSX}), which permits a
complete analysis of the most luminous infrared Galactic populations. The
images of this survey have also led to
the detection of more than 2000 infrared dark clouds (Egan et al.
\cite{msx-irdc}). A very recent analysis (Lumsden et al. \cite{msx-YSO})
of the MSX PSC has produced a large sample
of massive young stellar objects in the Galactic disk.

Among the many large observing programs conducted by ISO, including deep
and wide-field extragalactic surveys, worth mentioning are the European
Large-Area ISO Survey, ELAIS (Rowan-Robinson et al. \cite{ref-elais}),
ISOCAM deep surveys using guaranteed time observations (Elbaz et al.
\cite{ref-elbaz}), and FIRBACK, a deep 170~$\mu$m imaging survey carried out
with ISOPHOT (Dole et al. \cite{ref-firback}). Apart from these there were
also a number of observations of specific targets in the Galaxy. The following
ISOCAM
studies were with sensitivities comparable to or slightly deeper than ISOGAL (in
more limited areas): LW2 and LW3 imaging surveys of nearby star forming regions
(Nordh et al. \cite{nordh}; Bontemps et al. \cite{Bontemps}),
photometric studies of other Galactic \textsc{Hii}
regions (Zavagno \& Ducci \cite{zavagno}), and
the GPSURVEY (Burgdorf et al. \cite{GPsurvey}), which provided observations
of about 0.25~deg$^{2}$ in the central Galaxy at mid-infrared wavelengths.

In this paper, we give a detailed description of the ISOGAL observations
in Sect.~\ref{sec_obs}, and of their
processing and the related quality checks in Sect.~\ref{sec_data}.
The DENIS data are presented in Sect.~\ref{sec_denis}.
The content of the Point Source Catalogue (PSC) is explained
in Sect.~\ref{sec_psc}, and the complete descriptions of
various support tables are given in the relevant sections.
Finally, the main characteristics of the catalogue are briefly summarised
in Sect.~\ref{sec-summary}.

\section{\label{sec_obs}ISOGAL Observations and Fields}

\subsection{\label{sec-isog-obs}ISOGAL observations}

The mid-infrared observations were obtained with the ISO\-CAM instrument
(Cesarsky et al. \cite{ISOCAM}; Blommaert et al. \cite{isocam_hb})
on ISO (Kessler et al. \cite{ISO_sat}) using
filters centred at $\lambda \approx$~7 and 15~$\mu$m and with a pixel scale of
6'', or 3'' in a few cases. Table~\ref{filters} lists the filters used.
Most observations were performed with the broad filters LW2 and LW3, with a
field selection avoiding bright IRAS sources susceptible to detector array
saturation. However, a few regions with stronger sources
(around the Galactic Centre and in a few star forming regions) were
observed with the narrow filters LW5 or LW6, and LW9, and with smaller
pixel field of view (3'').

\begin{table}[htbp]\normalsize
\begin{minipage}{\linewidth}
\caption[]{\label{filters}ISOCAM filters used for ISOGAL: reference wavelengths
and bandwidths, zero point magnitudes and flux densities, and total observed
area.}
\begin{center}
\begin{tabular}{|c|c|c|cc|c|}
\hline
Filter & $\lambda_{ref}$ & $\Delta\lambda$ &
ZP\footnote{The magnitude of a source with a flux
density $F_{\nu}$ expressed in $mJy$ is given by 
$mag=ZP-2.5 \times log (F_{\nu}$)} & $F_{mag=0}$ & Area \\
& [$\mu m$] & [$\mu m$] & [mag] & [Jy] & [deg$^2$] \\
\hline
LW2 &  6.7 & 3.5 & 12.39 & 90.36 & 9.17 \\
LW5 &  6.8 & 0.5 & 12.28 & 81.66 & 0.64 \\
LW6 &  7.7 & 1.5 & 12.02 & 64.27 & 2.97 \\
\hline
LW3 & 14.3 & 6.0 & 10.74 & 19.77 & 9.92 \\
LW9 & 14.9 & 2.0 & 10.62 & 17.70 & 3.53 \\
\hline
\end{tabular}
\end{center}
\end{minipage}
\end{table}

For standard ISOGAL observations (broad filters LW2 and LW3), we estimated
that, to avoid saturation of the detector, no IRAS source with
F$_{\rm 12\mu m}$~$\geq$~6~Jy should be observed. This limit was further relaxed
up to F$_{\rm 12\mu m}$~$<$~20~Jy with narrow filters; however, even with such
a high limit value, it implied that a few regions, including the Galactic
Centre itself, could not be observed. A quick inspection of the images showed
that only very few observed pixels among all ISOGAL observations
were slightly above the limit of the linear
domain of the detector. The profiles of the associated
point sources do not deviate much from the average point spread function
(PSF, see Sect.~\ref{sec-psf-proc}), so that no source suffers
strongly from saturation in the published point source catalogue.

\begin{table*}[htbp]\normalsize
\begin{minipage}{\textwidth}
\caption{\label{ISO-obs}Format of ISOGAL Observations Table (version 1) - 384 entries
(see examples in Table~\ref{ISO-obs_ex})}
\begin{center}
\begin{tabular}{|rllll|} 
\hline
col & name & format & units [range] & description \\
\hline
 1 & ION       & a8   &                  & ISO Observation number  \\
 2 & name      & a13  &                  & ISOGAL observation name \\
 3 & date      & a6   & YYMMDD	         & date of observation \\
 4 & j\_day    & i4   &                  & Julian day of observation - 2450000\\
 5 & qual      & i1   & [1,2]            & quality of image\footnote{Image
quality: 1 is standard quality, 2 is medium quality (in most cases, the problem
is that the first individual image of the raster appears brighter than the other
ones). Images of bad quality have not been used to build the catalogue.}\\
 6 & l\_off    & f5.1 & arcsec           & applied offset in Galactic
longitude\footnote{The astrometry of the published raster images has been
corrected to match the DENIS astrometry if any (see Sect.~\ref{new-images}).
The offset values given in this table have been added to the initial raster
coordinates.}\\
 7 & b\_off    & f5.1 & arcsec           & applied offset in Galactic latitude \\
 8 & G\_lon    & f8.4 & deg [-180--+180] & Galactic longitude of raster centre \\
 9 & G\_lat    & f8.4 & deg [-90--+90]   & Galactic latitude of raster centre \\
10 & dl        & f6.4 & deg              & half width of raster in longitude \\
11 & db        & f6.4 & deg              & half width of raster in latitude\\
12 & RA        & f8.4 & deg              & RA (J2000) of raster centre \\
13 & DEC       & f8.4 & deg              & Dec (J2000) of raster centre \\
14 & filt      & i1   & [2,3,5,6,9]      & LW filter number  \\
15 & pfov      & i1   & arcsec [3,6]     & pixel field of view \\
16 & mag\_lim  & f5.2 & mag              & ISO magnitude cutoff\footnote{The ISO
magnitude cutoff has been computed for each observation to correspond at least
approximately to a $50\%$ completeness level (see Sect.~\ref{artif}).} \\
17 & nb\_sour  & i4   &                  & number of extracted sources brighter
than mag\_lim \\
18 & rot       & i1   & [0,1]            & applied transformation (270$^{\circ}$
rotation) to the raster\footnote{Columns 18, 19 and 20: all the published images
are oriented with $l$ along decreasing $x$ and $b$ along increasing $y$. In each
column, a 1 means that the corresponding transformation has been applied to the
initial (OLP7 processed) raster, and a 0 means that this transformation was not
needed.} \\
19 & x\_inv    & i1   & [0,1]            & applied transformation (x-inversion)
to the raster \\
20 & y\_inv    & i1   & [0,1]            & applied transformation (y-inversion)
to the raster \\
21 & m         & i2   &                  & number of raster steps in x in final raster \\
22 & n         & i2   &                  & number of raster steps in y in final raster \\
23 & dm        & i3   & arcsec           & size of step between x (final) raster positions \\
24 & dn        & i3   & arcsec           & size of step between y (final) raster positions \\
25 & angle     & f6.2 & deg              & angle from the upward axis to the north
in the final raster \\
26 & NX        & i3   & pixel            & number of pixels in x of final raster \\
27 & NY        & i3   & pixel            & number of pixels in y of final raster \\
\hline
\end{tabular}
\end{center}
\end{minipage}
\end{table*}

The observations were performed as rasters. The basic ISOCAM observation
is a $32\times32$ pixel image of 0.28 sec integration time. Due to
limitations in the downlink data rate, these basic images were coadded in
groups of four and downlinked, making the unit frame one of 1.12 sec
integration time. At each raster position 19 such frames were obtained,
resulting in an integration time of $\sim 21\,$sec per raster position.
The rasters were oriented along galactic latitude and longitude, which
differed from the direction of the sides of the detector array, resulting in
``saw-tooth'' edges of the final mosaics. With 6'' pixels, the raster steps
were typically 90'' in one direction and 150'' in the perpendicular one
(and a factor of two smaller with 3'' pixels), in order to observe
each sky position about twice. However, because of the non-alignment of the raster
and detector axes, each sky position was not as regularly observed.
The actual number of observations per sky point varied from four to
exceptionally zero (for the dead ISOCAM column close to a raster
edge), with an average of $\sim$~1.5.

The total area covered by the ISOGAL survey is $\sim$~15.6 square degrees,
of which 10.7 were observed at both 7 and 15~$\mu$m, 2.1 were observed at
7~$\mu$m only, and 2.8 were observed at 15~$\mu$m only. This survey is
the result of three successive proposals developed over the lifetime of
ISO.
As a consequence, most fields were observed at
7 and 15~$\mu$m at different dates, and some fields were
observed at one wavelength only, in particular because the planned
targets were not observable at the very end of the mission.

A total of 696 observations compose the ISOGAL survey. Of all these
observations, 29 could not be used because of instrument failures or other
problems during the data reduction. Another 18 observations are single
ISOCAM frames (32x32 pixels) observed in the spectroscopic
{\em Circular Variable Filter} (CVF) mode; they are treated in a
different way (Blommaert et al., in preparation). A further 186 images are
``dummy'' observations, containing only one $32\times32$ pixel image
- acquired after repositioning of the telescope to allow for
reconfiguring the camera from the CAM parallel mode to that of
the observation - and have not been used for the catalogue. As a result, only
463 raster-observations are considered as relevant for the imaging survey.

\begin{figure}[htbp]
\begin{center}
\resizebox{8.8cm}{!}{ \includegraphics{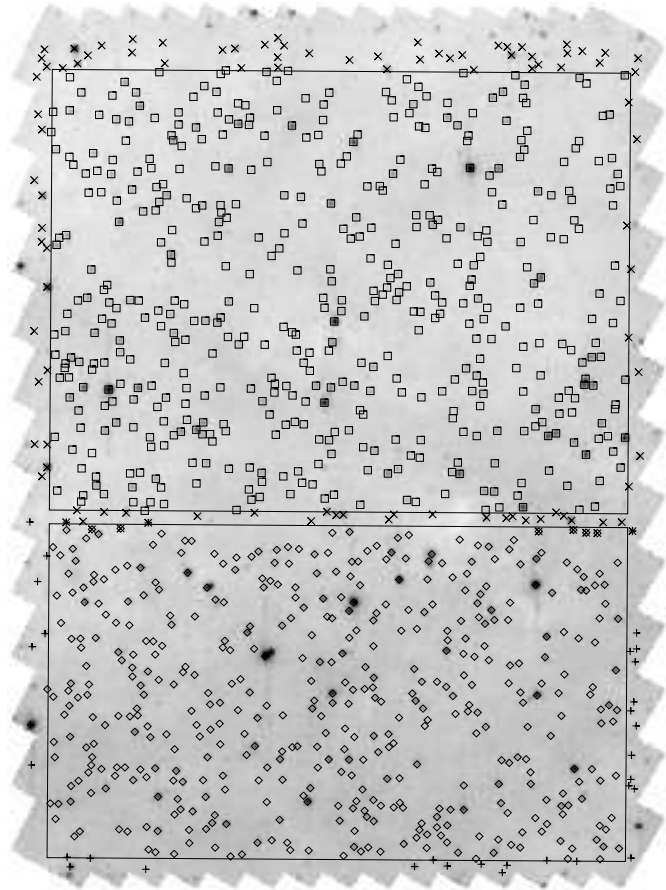}}
\caption[]{Example of one ISOGAL observation which has been used for 
one FA and one FC fields. The formal limits of both fields are shown
with rectangular frames: FC field (upper frame) and FA field (lower
frame). The different symbols correspond to the different catalogues
of sources (see Sect.~\ref{sec_psc}): squares (FC, regular), crosses
(FC, edge), diamonds (FA, regular) and plus signs (FA, edge).}
\label{field_fafc}
\end{center}
\end{figure}

To avoid redundancy in the published catalogue (due e.g. to various
observations of a test field with several filters, but also to small
overlapping areas between two observations in many cases), we decided to
use, for the present version of the PSC, only one observation at 7~$\mu$m
and one at 15~$\mu$m for a given position\footnote{
However, in very few cases due to edge effects, two ISOGAL sources have
exactly the same final coordinates because they are associated with
the same DENIS source (see also Sect. \ref{sec-psc-1})}.
Thus, we had to choose the best observation in the case of
overlapping images at the same wavelength. The selection criteria were:
first, if the different observations are obviously of different quality,
the best quality one was selected. Then, if the observations were made
with different filters, we chose to keep the one with a broad filter (if
it exists) because the number of detected sources is larger. In the very
few cases where the filter
is the same but the pixel size is different, we selected the large (6'')
pixel observations in order to have more homogeneous data. If the quality
and the observational setup were approximately the same in different
observations, we then selected the most recent one (the one with higher
ISO observation  number), because on average the data quality was better
certified. Finally, 384 raster images have been used to build the PSC.

All the raster images used are published with the PSC (and available through
the CDS and IAP web sites\footnote{
http://www-isogal.iap.fr/Fields/index\_tdt.html}), and the electronic
version of the Catalogue of ISOGAL Observations of the PSC contains
384 entries, each entry having the format described in Table~\ref{ISO-obs}.
Two examples are shown in Table~\ref{ISO-obs_ex}, for the 7 and 15~$\mu$m
observations composing a test field of 0.027~deg$^2$ centred at
$(l,b) = (0.0,1.0)$, hereafter called the ``C32'' field.


\begin{table*}[htbp]\normalsize
\caption{\label{ISO-obs_ex}Two examples of entry in the ISOGAL Observations Table
(see Table~\ref{ISO-obs} for explanation),
from the ``C32'' field at $(l,b) = (0.0,1.0)$}
\begin{tabular}{|l|cccc@{ }cc@{ }cccccc|}
\hline
col & 1 & 2 & 3 & 4 & 5 & 6 & 7 & 8 & 9 & 10 & 11 & 12 \\
name & ION & name & date & j\_day & qual & l\_off & b\_off & G\_lon & G\_lat &
dl & db & RA \\
\hline
Ex. 1 & 83600418 & 2P00P10B & 980228 & 873 & 1 & -4.8 & -5.6 & 0.0001 & 0.9988 &
0.1633 & 0.0758 & 265.4353 \\
Ex. 2 & 83600523 & 3P00P10B & 980228 & 873 & 1 & -6.3 & -3.1 & -0.0003 & 0.9995 &
0.1633 & 0.0758 & 265.4353 \\
\hline
\end{tabular}
\vspace{1 mm}
\begin{tabular}{|l|cccccccccc@{ }c@{ }cccc|}
\hline
col  & 13 & 14 & 15 & 16 & 17 & 18 & 19 & 20 & 21 & 22 & 23 & 24 & 25 & 26 & 27 \\
name & DEC & filt & pfov & n & m & dn & dm & mag\_max & nb\_sour & rot &
x\_inv & y\_inv & angle & NX & NY \\
\hline
Ex. 1 & -28.4136 & 2 & 6 & 7 & 4 & 150 & 90 & 8.89 & 331 & 1 & 0 & 0 & 58.95 & 196 & 91 \\
Ex. 2 & -28.4136 & 3 & 6 & 7 & 4 & 150 & 90 & 8.00 & 220 & 1 & 0 & 0 & 58.97 & 196 & 91 \\
\hline
\end{tabular}
\end{table*}

\subsection{Definition and list of ``Catalogue Fields''}

We define an ISOGAL ``field'' as a rectangular area of the sky whose edges
are aligned with the galactic axes, and which has been completely observed
with ISOCAM. There are three kinds of fields, depending on the available
observations: the ``FA'' fields were observed only at 7~$\mu$m, the
``FB'' fields were observed only at 15~$\mu$m, and the ``FC'' fields
were observed at both 7~$\mu$m and 15~$\mu$m.

\begin{table*}[htbp]\normalsize
\renewcommand{\thefootnote}{\textit{\alph{footnote}}}
\begin{minipage}{\textwidth}
\caption{\label{fields-cat}Format of ISOGAL ``Fields'' Table (version 1) - 263
entries (see example in Table~\ref{fields-cat_ex})}
\begin{center}\begin{tabular}{|rllll|}
\hline
col & name & format & units [range] & description \\
\hline
 1 & Name      & a14  &                  & ISOGAL field identifier \\
 2 & ION7      & a8   &                  & ION for 7~$\mu$m data (see
Table~\ref{ISO-obs}) \\
 3 & ION15     & a8   &                  & ION for 15~$\mu$m data \\
 4 & filt7     & i1   & [2,5,6]          & 7~$\mu$m filter \\
 5 & filt15    & i1   & [3,9]            & 15~$\mu$m filter \\
 6 & pfov      & i1   & arcsec  [3,6]    & pixel field of view \\
 7 & G\_lon    & f8.4 & deg [-180--+180] & Galactic longitude of field centre \\
 8 & G\_lat    & f8.4 & deg [-90--+90]   & Galactic latitude of field centre \\
 9 & dl        & f6.4 & deg              & half width of field in
longitude\footnote{$dl$ and $db$ apply to the limits inside the edges of
the images within which sources are accepted.} \\
10 & db        & f6.4 & deg              & half width of field in
latitude\footnotemark[1]{} \\
11 & area      & f6.4 & deg$^2$          & area of field \\
12 & dens7     & i5   & deg$^{-2}$       & density of 7~$\mu$m sources \\
13 & dens15    & i5   & deg$^{-2}$       & density of 15~$\mu$m sources \\
14 & RMS\_II   & f4.2 & arcsec           & RMS separation of 7-15~$\mu$m associated
sources \\
15 & RMS\_ID   & f4.2 & arcsec           & RMS separation of ISO-DENIS associated
sources \\ 
16 & K\_max1   & f4.1 & mag              & DENIS $K_{\rm s}$ magnitude cutoff
1\footnote{maximum DENIS $K_{\rm s}$ magnitude limiting the density of $K_{\rm s}$ DENIS
sources to $\sim$~18~000 sources per square degree if the ISO images have 6''
pixels (or to $\sim$~72~000 sources per square degree for the 3'' ISO
observations). K\_max1 is used to discuss the quality of ISOGAL--DENIS
associations (see Sect.~\ref{sec-flagID}).} \\
17 & K\_max2   & f4.1 & mag              & DENIS $K_{\rm s}$ magnitude cutoff
2\footnote{maximum DENIS $K_{\rm s}$ magnitude accepted in order to avoid spurious
cross-identifications. The density of $K_{\rm s}$ DENIS sources is limited to
$\sim$~36~000 sources per square degree for 6'' ISO observations (and again
to $\sim$~72~000 sources per square degree for 3'' ISO observations).} \\
18 & dens\_K2  & i5   & deg$^{-2}$       & density of DENIS $K_{\rm s}$ sources
used\footnote{density of DENIS $K_{\rm s}$-band sources brighter than the cutoff
magnitude K\_max2.} \\
\hline
\end{tabular}
\end{center}
\end{minipage}
\end{table*}

To build the present version of the PSC, we have defined a total of
43 FA fields, 57 FB fields and 163 FC fields.
In some cases, a fraction of an ISOGAL observation was
used for an FA (or FB) field, and another fraction was used for an FC field
(see e.g. Fig.~\ref{field_fafc}),
so that only 384 different observations were required for these 263 fields.
These peculiar configurations can result in the presence of a few redundant
sources: because of edge effects, two sources at the same position may
appear in two different catalogues; nine such cases can be seen on
Fig.~\ref{field_fafc} (see also Sect. \ref{sec-psc-1}).
The complete catalogue of the 263 ISOGAL fields is available
electronically\footnote{http://www-isogal.iap.fr/Fields/ and from the VizieR
service at CDS: http://vizier.u-strasbg.fr/viz-bin/VizieR} and contains 18
columns, as described in Table~\ref{fields-cat}, and an example is given
in Table~\ref{fields-cat_ex}.

\begin{figure*}[htbp]
\begin{center}
\resizebox{17.3cm}{!}{ \rotatebox{90}{\includegraphics{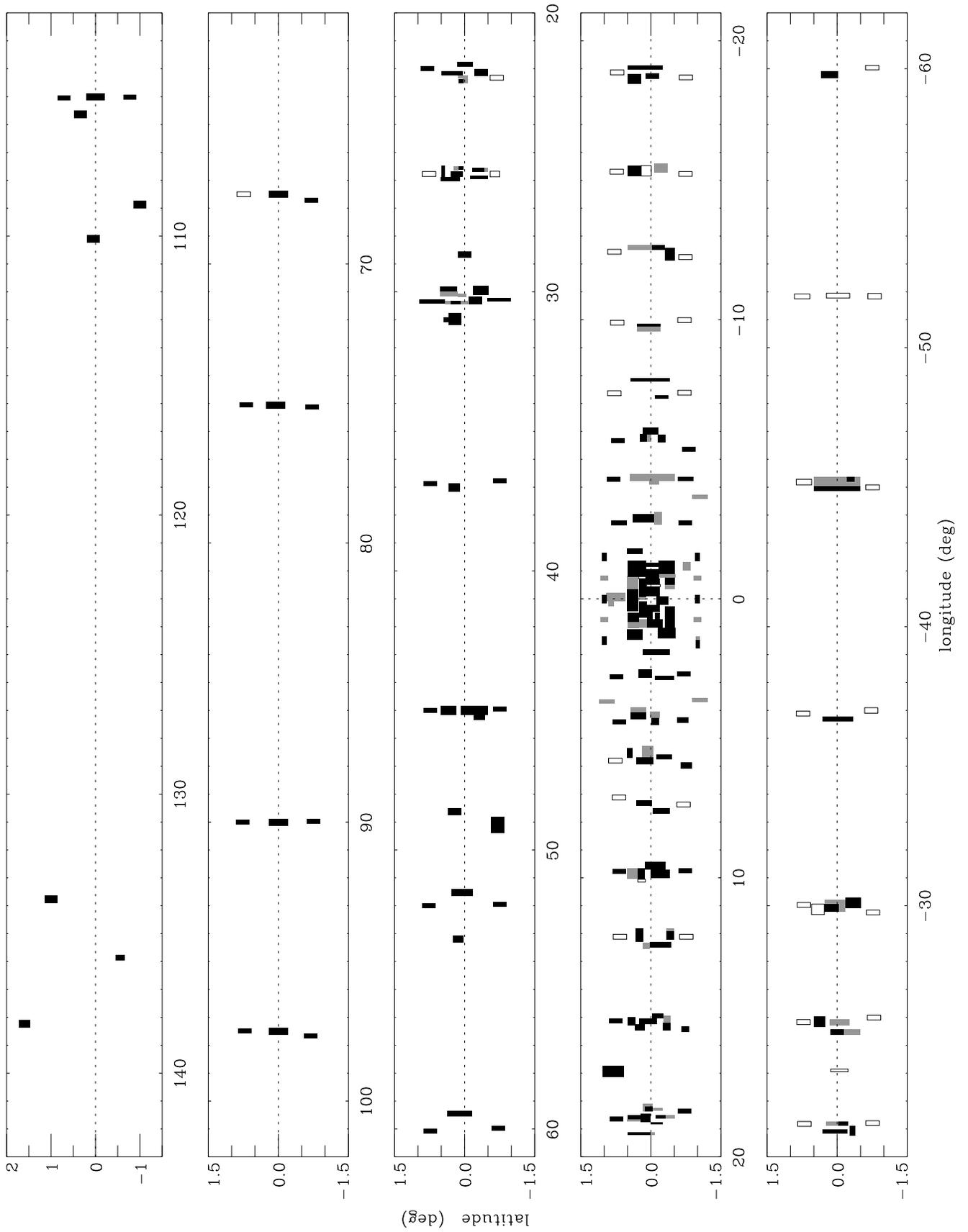}}}
\caption[]{Galactic map of the ISOGAL disk fields. The black boxes show
the fields which have been observed at both 7 and 15~$\mu$m (FC fields), while
the open boxes stand for 7~$\mu$m only observations (FA), and the
grey ones for 15~$\mu$m only observations (FB).}
\label{dfields}
\end{center}
\end{figure*}

The field names are generated using 14
characters, and the first two indicate the type of the field (FA, FB or FC).
The 12 last characters of the field names are the
galactic coordinates in decimal degrees of the centre of the field.
A graphical view of the observed fields is given in Fig.~\ref{dfields}.

\begin{table}[htbp]\normalsize
\caption{\label{fields-cat_ex}Example of entry in the ISOGAL Fields Table
(Table~\ref{fields-cat}) (``C32'' field at $(l,b) = (0.0,1.0)$)}
\begin{center}\begin{tabular}{|rl|c|} 
\hline
col & name & C32 field \\
\hline
 1 & Name      & FC+00000+00100 \\
 2 & ION7      & 83600418 \\
 3 & ION15     & 83600523 \\
 4 & filt7     & 2 \\
 5 & filt15    & 3 \\
 6 & pfov      & 6 \\
 7 & G\_lon    & -0.0011 \\
 8 & G\_lat    & 0.9990 \\
 9 & dl        & 0.1441 \\
10 & db        & 0.0471 \\
11 & area      & 0.0271 \\
12 & dens7     & 9225 \\
13 & dens15    & 6125 \\
14 & RMS\_II   & 2.24 \\
15 & RMS\_ID   & 1.70 \\
16 & K\_max1   & 9.6 \\
17 & K\_max2   & 10.6 \\
18 & dens\_K2  & 35979 \\
\hline
\end{tabular}
\end{center}
\end{table}

\section{\label{sec_data}ISOGAL data processing and quality}

A complete description of the data processing and of the procedures that were
run to quantify the quality of the data is given in the electronic version of
this paper, available through the ADS service\footnote{Or directly from the
A\&A editor's web site,\\
http://www.edpsciences.org/journal/index.cfm?edpsname=aa}.
Only the main results,
which may be useful to all users of the catalogue, are summarised in this
section and the following one.

\subsection{\label{sec-isocam}ISOCAM image processing}

Data reduction was performed with standard procedures of
the CAM Interactive Analysis
(CIA, Ott et al. \cite{ref-CIA}) package version 3.0 on data products
produced with version 7.0 of the ISO Off-Line Processing (OLP) pipeline
(Blommaert et al. \cite{isocam_hb}). A particular care was taken to
correct the effect of the slow detector responsitivity.
Two methods of transient correction
were used: the IAS model transient correction (Abergel et al. \cite{Abergel}),
also called the `inversion' method, was used to get the best photometry for
non-stabilised signals, while the auxiliary `vision' method
(Starck \cite{ref-vision};
Starck et al. \cite{ref-vision2}) was used to remove most of the latent
images (or remnants) due to memory effects of the detectors to strong sources
(Coulais \& Abergel \cite{Coulais}). We thus have two sets of reduced data:
a) the main one treated with `inversion', which performs a
correction for the missing signal, (though this correction is not perfect, see
Sect.~\ref{sec-trans-corr}), but which still contains the remnants;
b) an auxiliary one, roughly treated with `vision', where most remnants
have been removed, but with wrong photometry.
The two rasters are converted to physical units (mJy),
using the standard conversion factors (Blommaert \cite{photom_rep}).

\subsection{\label{extract}Point source extraction}

\subsubsection{\label{sec-psf-proc}The point source extraction procedure}

A dedicated PSF fitting procedure worked out by C. Alard has been used to
extract point sources from all `inversion' and `vision' processed images.
First, a search for local maxima is performed on the complete image,
resulting in a list of pixel positions of point source candidates. Then, an
analytical expression of the PSF is fitted at each position to compute the
flux density of the point sources, and to discard the local maxima whose
shapes do not correspond to the instrumental response to a point source.


The source detection is first performed on an oversampled image,
using pixels a factor of two smaller than in the initial image.
This image is used only for the detection step
of the source extraction. The oversampling is performed by a
convolution of the initial pixels with an analytical expression of
a theoretical PSF. As a result, the sources can be localised on
a thinner grid.

The detection procedure looks for local maxima in the oversampled image.
This step is controlled by a {\em mesh} parameter, which can take
values of 1 or 2, and defines the size of the grid on which local maxima are
looked for. With $mesh=1$, all local maxima are detected, even those
corresponding to bright spots in the background rather than to point sources.
On the other hand, with $mesh=2$, 5$\times$5 oversampled pixel meshes are
used to find local maxima, resulting in a smoothing of the irregularities in
the background, without any significant loss in the detection of relatively
bright ($F_{\nu} \ga$ 100~mJy) point sources, but with a more confusion
limited extraction of the faintest sources.

The extraction procedure which has been used to build the ISOGAL PSC
performed a complete extraction with each value of $mesh$. For each
observation, the two resulting catalogues
have been cross associated to check the quality and the reality of the detected
sources (see Sect.~\ref{extrac-assoc}). Obviously the extraction performed with
$mesh=1$ is the most efficient to correctly extract blended sources; on the other
hand, a non negligible fraction of the sources extracted only with $mesh=1$
(with no association in the extraction performed with $mesh=2$) seem to
be spurious (see the discussion in Sect.~\ref{sec-reality}).


Another procedure is used to measure the flux density of the sources
on the original image, and to estimate the correlation of their profile
with the PSF. For each observational setup (combination of one filter
and one pixel scale), a single reference PSF has been determined for all
the observations from a sample of relatively bright and isolated sources.
A least square fit between the reference profile and a 5$\times$5 (not
oversampled) pixel mesh is computed at the position of each source candidate,
starting with the brightest one. The background
is estimated from the median value of the pixels in an
annulus of inner and outer radii equal to 3 and 5 pixels, respectively.
The results of this operation are the flux density of the source and the
uncertainty on its measurement, computed as the RMS of the residual between the
scaled PSF profile and the actual source profile. This flux density uncertainty
is later converted to a magnitude uncertainty, hereafter called $\sigma$.
The reality of each
point source is estimated by the ratio of the fitted flux density to the
RMS uncertainty of the fit, and only sources with this ratio greater than 3
are considered valid and stored in the resulting catalogue. Then, the
profile of the source is subtracted from the image, and the procedure
runs iteratively going to fainter and fainter sources. This method is
powerful even in crowded fields, where it is able to estimate correctly
the flux densities of blended sources.

\subsubsection{\label{extrac-assoc}Source quality checks}

Four catalogues have been built for each observation, combining the
two possible values of $mesh$ (1 or 2) and the `inversion' and `vision'
processed rasters. Considering the high background level in the Galactic
Disk, we decided to anyhow limit the published catalogue to a flux density
of 5~mJy ([7]$\approx$10.5 and [15]$\approx$9.0) to reduce the number of
spurious sources (another limit was eventually later applied depending
on the field, see Sect.~\ref{artif}).
Then, the sources found in `inversion' processed images that were associated
with a `vision' source within a search radius of one observed pixel
were considered valid, while those found
only in the `inversion' images were considered spurious (these can be
remnants of bright sources, or other non real point-like sources).
The distance between the `inversion' and the `vision' sources gives a good
estimate of the quality of the sources: it is generally smaller than 1'' for
real sources, while a separation larger than 3'' may be due to artifacts
(see also Sect.~\ref{sec-spurious}).
The final data (position and photometry) in the catalogue come only from
the `inversion' processed rasters, with elimination of the remnant sources
using the `vision' results.

The majority (70\%) of the extracted sources could be associated between the
$mesh=1$ and the $mesh=2$ catalogues (with a 6'' association radius for all
observations), while the remaining 30\% are only found with $mesh=1$.
Since less than 1\% of the
extracted sources were detected with $mesh=2$ with no counterpart in the
$mesh=1$ catalogue, while almost 30\% of the extracted sources were only
detected with $mesh=1$, the published data (position and photometry) come
from the $mesh=$1 results for the sources which were detected with both
values, in order to get a homogeneous set of data. Further quality selection
criteria are applied later in the processing (see Sect.~\ref{sec-spurious}),
so that only $\approx$10\% of the sources in the published catalogue
have been detected with $mesh=$1 only.
A special MESH flag is included in the catalogue to indicate for which
value(s) of $mesh$ a source has been extracted, and the global QUALITY
flag is decreased for sources without association between the
$mesh=1$ and the $mesh=2$ results (see next Section).

\subsubsection{\label{psc-mesh}Source extraction quality flags}

The quality of the derived photometry as well as the reliability of the
extracted sources can be affected by several factors, and different quality
flags have been computed to warn the user when effects degrading the
photometric quality are present, and to finally estimate the global
quality of the point sources.

\subsubsection*{The MESH flag}
The MESH flag is set to 1 (resp. 2) for
sources which have been detected only with $mesh=1$ (resp. 2), and to 3 for
the sources which could be associated between the two extractions
(see Sect.~\ref{extrac-assoc}), thus making their reality more trustful.

\subsubsection*{The NPIX flag}
The number of independent measurements of the signal at the position
of a source, which takes into account the number of coadded individual
exposures, but also the fact that some exposures might be discarded due
to glitches or to the ISOCAM dead column, directly
affects the photometric quality. The NPIX flag is the integer part of one
tenth of the weighted number of measurements usable at the central
position of the source, as given in the third plane of the OLP7 processed
FITS files. As each raster position has been observed
on average 19$\times$1.5 (see Sect.~\ref{sec-isog-obs}) times,
typical ``good quality'' values of this flag are in the range 2 to 4. Note
that this flag is rather an indication of the number of good exposures than
a number of pixels involved, but we decided to keep the NPIX name, as it
appears in the header of the OLP7 processed files.

\subsubsection*{The EDGE flag}
The position of a source with respect to the edges of the raster also
affects the derived photometric quality, because the extraction procedure
needs a large enough observed area to properly compute the flux density
of the source and the background to be subtracted. The
EDGE flag is set to 1 when the centre of the source is at a distance
between two and five pixels from the edge of the observed raster (taking
into account the saw-tooth borders), and to 0 when the distance is greater
than five pixels. Sources at less than two pixels from one edge were removed
from the catalogue, since their flux density cannot be properly estimated.

\subsubsection*{\label{psc-quality}The global quality flag Q}

\begin{figure}[htbp]
\begin{center}
\resizebox{8.5cm}{!}{ \rotatebox{0}{\includegraphics{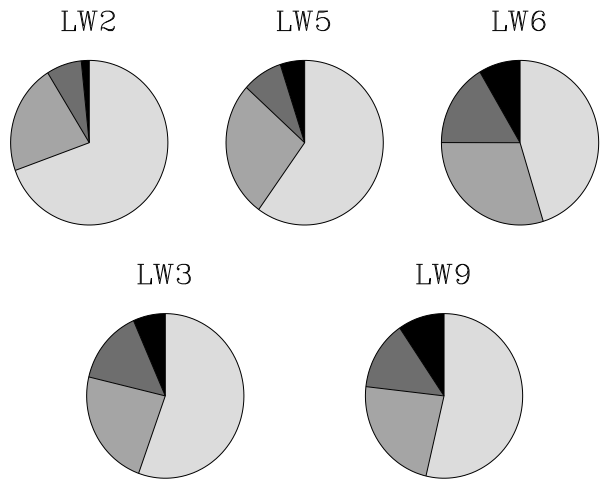}}}
\caption{Distribution of the quality flag Q for the different filters. The gray
scale corresponds to the different values of this flag, from 4
(lightest grey) to 1 (darkest grey).}
\label{pie_Q}
\end{center}
\end{figure}

A global quality flag Q was computed by combining the flags defined above
and the photometric uncertainty $\sigma$. Its value ranges from 1 to 3 for
sources with MESH $=$ 1 or 2, and from 2 to 4 for sources with MESH
$=$ 3, the higher the better quality.
%
%
The distribution of this flag for all the sources in the
catalogue is shown for the different filters in Fig.~\ref{pie_Q}. As can be
seen, more than one half of the sources have a very good photometric quality
(Q = 4). A value of 3 for this flag can also be considered as reasonably good
quality. Finally, only $\sim 15$\% of the sources in the catalogue have a
moderate photometric quality (Q $\leq$ 2). They should be used with much
caution since their reliability is not warranted.

Additional estimates of the reliability of the sources are provided by
the analysis of repeated or overlapping observations (see
Sect.~\ref{sec-repeated}), but also by the combination of several wavelengths,
including DENIS ones:
a source with a moderate quality flag at, for example, 7~$\mu$m, but with
a good quality association at 15~$\mu$m (see Sect.~\ref{sec-715-assoc})
finally has a very large probability to be a real source.

\subsubsection{\label{extended-src}Extended sources}

The first version of the ISOGAL PSC only contains point sources, and
sources of very small extension. The extraction of extended objects
will be performed with a dedicated procedure for the second version
of the catalogue.

The present version of the PSC contains a small proportion of
sources of small extension, with typical sizes
around 10-20'' (FWHM). These slightly-extended sources are characterised
by relatively high values of the photometric uncertainty, with typical
$\sigma \approx$ 0.15 mag for bright ($F_{\nu} \approx$ 1~Jy) sources,
while bright point sources generally have $\sigma < 0.05$ mag.
Aperture photometry performed on a small sample of such bright slightly
extended sources has shown that their magnitudes can be underestimated
by about 1 mag (Schuller \cite{phd-schuller}). It is planned to perform
accurate photometry and to
include a relevant extension flag in the second version of the PSC.

\subsubsection{\label{sec-spurious}Spurious sources}
Three kinds of extracted sources are considered as spurious: (1) the
``inversion-only'' sources, i.e. those found in `inversion' rasters
with no counterpart in the `vision' rasters,
(see Sect.~\ref{extrac-assoc}), (2) the
sources with an inversion-vision association with a large separation
($\geq$0.5 pixel) and with a poor extraction confirmation
(flag $MESH < 3$),
and (3) the other possible remnants of bright sources.
Indeed, the `vision' method (see Sect.~\ref{sec-isocam})
does not remove all remnant sources, and remaining
remnants of bright ($\geq$~100~mJy) sources were identified 
by looking for faint sources within a radius of 0.5 pixel
around the exact location of the bright source in the detector
at the five successive positions in the raster.
They have been removed from the catalogue
and their positions and magnitudes are listed
in the catalogue of spurious sources (Sect.~\ref{spurious}). Unfortunately,
true faint sources which are found at the position of a putative remnant are
also considered as spurious, and appear in the catalogue of spurious sources
but not in the PSC. 

\subsection{\label{sec-trans-corr}Photometric calibration}

The flux densities of the point sources, as obtained by the PSF fitting
procedure, lead to a good relative photometry, but have to be calibrated
in an absolute way. Two factors introduce biases in the photometry.
First, the integration time for the standard ISOGAL observations was
too short to allow the signal to stabilise.
A correction to this transient problem is applied with the
`inversion' method (see Sect.~\ref{sec-isocam}). However, this method only
allows proper correction for extended emission, but is insufficient
for point sources (Coulais \& Abergel \cite{Coulais},
Blommaert \cite{photom_rep}).
A few ISOGAL fields were observed with longer integration times. A comparison
between {\em regular} and {\em long} measurements showed that the photometry
from the {\em regular} raster is about 0.2 mag too high (too faint).

Our PSF photometry introduces a second bias, because a fraction of the signal
from a point source falls outside the mesh used to model the PSF.
The general flux calibration of ISOCAM was established from measurements on
standard stars (Blommaert \cite{photom_rep}; Blommaert et al.
\cite{Blom2000}). The observed signal was measured using aperture photometry,
which was corrected for the part of the PSF falling outside the aperture.
To convert our PSF-fitting photometry to absolute photometry, a comparison
was made with photometry obtained using the same techniques as in the ISOCAM general
flux calibration.
The aperture magnitudes were found to be lower (brighter) than the
PSF magnitudes by 0.2--0.4~mag, revealing a bias in the PSF
normalisation.

\subsubsection{Final correction}

The total correction
that has to be applied is between $-0.37$ and $-0.59$ mag for the different
setups. As the uncertainty on each determined correction is at least 0.1
mag we decided to apply the same constant offset of $-0.45$ mag to all
the sources and for all observational setups. This correction leads to
photometry in good agreement with external comparison data, as is
explained below.

The first publications based on ISOGAL data made use of a non-corrected
photometry. The mid-infrared magnitudes presented there should thus
be corrected by a $-0.45$ mag offset (with a possible $\pm$~0.1
mag additional discrepancy from field to field). This concerns in particular
the results published in P\'erault et al. (\cite{perault}), Testi et al.
(\cite{testi97}), Omont et al. (\cite{omont99}), Glass et al. (\cite{glass99}),
Schultheis et al. (\cite{variable}) and Felli et al. (\cite{felli00}).
Appropriate errata will be published for these papers.

\subsubsection{External checks}

The comparison of the observed with the predicted photometry for stars
with known spectral types and distances provides an absolute calibration.
Comparing the predicted and the corrected PSF magnitudes for three stars
from the Hipparcos Input Catalogue we obtain:
\[ mag_{\rm pred} - mag_{\rm PSF} = 0.04 \pm 0.10 \]
where the result is the
average of all determinations, independent of filter-PFOV combination,
and the uncertainty is the variance of the six determinations obtained,
though the distribution of these determinations is clearly non-Gaussian. 

A second check on the photometry is provided by the cross calibration with
the published catalogue of bright sources detected by the MSX
survey of the Galactic Plane (Price et al. \cite{ref-MSX}).
A comparison with the Band D photometry of MSX, which
used a filter similar to the ISOCAM 15~$\mu$m filters, showed good agreement
between the corrected ISO magnitudes and the MSX ones. For 650 stars
(424 observed with LW3 and 226 with LW9) we find:
\[ mag_{\rm MSX} - mag_{\rm PSF} = 0.01 \pm 0.40 \]
where the uncertainty is the RMS of the measured differences in magnitude.
The large width of the distribution is due to the combination of
the ISO and MSX photometry uncertainties, and to the intrinsic variability
of many of such bright stars. Note that,
strictly speaking, this result is
valid for the brightest ISOGAL stars that could also be measured by MSX
(which means roughly [15]~$\la$~4.0). Moreover, the computation of the
mean difference in magnitudes was limited to an even brighter
sample ([15]~$<$~3.0) in order to avoid Malmquist bias.
This nevertheless shows that our photometric calibration is reasonably
good and in agreement with others.

\subsection{\label{artif}Artificial sources}

Artificial star experiments (see Bellazzini et al. \cite{ref-artif} and
references therein for a general description) were conducted on the ISOGAL
images in order to study the effects of a crowded field on the photometric
quality and the completeness of the extracted point source catalogue.
A procedure was created for adding artificial stars to the ISOGAL
images, for extracting the sources with the same pipeline as the
one used to generate the ISOGAL catalogue, and for checking how well
the input sources are extracted.


Artificial star experiments enabled us to evaluate both random and
systematic photometric errors due to crowding, as well as the completeness
level of the extraction.
The output magnitudes were found brighter than the input ones. This bias is
very small for bright stars, but can reach 0.3 magnitude for the faintest
ones in the densest fields, where the probability of blending with real
stars is higher (see e.g. Fig. E-10 in the electronic version of this paper).

The completeness of the extraction can be quantified as follows.
For each observation, we can plot the
fraction of simulated sources which were retrieved as a function of
input magnitude. We observe a smooth curve which drops for the
faintest magnitudes. The magnitude where this fraction becomes
less than 50\% strongly depends on the density of the field.
We used this trend to define the limiting magnitudes for each observation,
corresponding to the faintest sources that were included in the published
catalogue.
These magnitudes are generally consistent with the magnitudes above
which the bias reaches 0.1 magnitude and its standard deviation reaches 0.3
magnitude. We derived relations between source density and limiting magnitudes
for the different observational setups (see also Fig. E-13 and E-14 in the
electronic version). We make a distinction between the
core of the ISOGAL survey observed with broad filters and 6'' pixels
and the peculiar observations of difficult fields observed with
narrow filters and 6'' or 3'' pixels.

\subsubsection*{A) 6'' pixel observations with broad filters}

For the 6'' pixel observations with LW2 and LW3 filters,
we computed the following linear relations:
\begin{itemize}
\item
for LW2 observations:
\begin{equation}
\label{mag_lim6_1}
mag_{\rm lim} = \left\{ \begin{array}{ll}
10.1 & \mbox{if $d \leq 0.01$},\\
10.7 - 60. \times d & \mbox{if $d \geq 0.01$},
\end{array} \right.
\end{equation}
where $d$ is the source density expressed in source/pixel.
Thus, the limiting magnitude ranges from 10.1 to 8.8, corresponding to limiting
flux densities between 8 and 27~mJy.
\item
for LW3 observations:
\begin{equation}
\label{mag_lim6_2}
mag_{\rm lim} = \left\{ \begin{array}{ll}
8.7 & \mbox{if $d \leq 0.005$},\\
8.9 - 40. \times d & \mbox{if $d \geq 0.005$},
\end{array} \right.
\end{equation}
Here, the limiting magnitude ranges from 8.7 to 7.7, and the associated flux density
ranges from 6.5 to 16~mJy.
\end{itemize}

\subsubsection*{B) 6'' pixel observations with narrow filters}

The results of our artificial source simulations show that the completeness
level is generally worse in LW5, LW6 and LW9 observations, which can
be interpreted as an effect of the much brighter diffuse background in the
peculiar regions which needed the use of such narrow filters. Therefore,
we applied 0.5 magnitude brighter cutting criteria for the 6'' observations
with these filters:
\begin{itemize}
\item
for LW5 and LW6 observations:
\begin{equation}
\label{mag_lim6_3}
mag_{\rm lim} = \left\{ \begin{array}{ll}
9.6 & \mbox{if $d \leq 0.01$},\\
10.2 - 60. \times d & \mbox{if $d \geq 0.01$},
\end{array} \right.
\end{equation}
\item
for LW9 observations:
\begin{equation}
\label{mag_lim6_4}
mag_{\rm lim} = \left\{ \begin{array}{ll}
8.2 & \mbox{if $d \leq 0.005$},\\
8.4 - 40. \times d & \mbox{if $d \geq 0.005$},
\end{array} \right.
\end{equation}
\end{itemize}
In addition, the photometry of the faintest sources in these peculiar fields
is less accurate than in standard observations. Therefore we decided
to decrease the quality flags (see Sect.~\ref{psc-mesh}) for the sources 
with magnitudes between mag$_{\rm lim}$-0.5 and mag$_{\rm lim}$, and
we extended the range in which we decreased the quality flags down to
mag$_{\rm lim}$-1 for the most difficult
FC+01694+00081 field located in the M16 nebula.

\subsubsection*{C) 3'' pixel observations}

The situation is more complicated for the 3'' pixel observations, because
they are too few and peculiar to allow a global statistical treatment.
Artificial source simulations have been run on all the 3''
pixel observations used in the PSC,
and the
results show good agreement between the different observations with a given
filter. Therefore we used a single limiting magnitude for each filter, and the
different values are given in Table~\ref{tab-mag_lim3}.
These limits give reasonably good results in terms of bias and completeness.

\begin{table}[htbp]
\caption[]{\label{tab-mag_lim3}Limiting magnitudes used to cut the catalogues
for 3'' pixel observations.}
\begin{center}
\begin{tabular}{|l|c|c|c|c|}
\hline
Filter& LW2 & LW5 & LW3 & LW9 \\
\hline
mag$_{\rm lim}$ & 10.0 & 8.4 & 8.5 & 7.0 \\
\hline
\end{tabular}
\end{center}
\end{table}


\subsubsection{\label{artif_ccl}Conclusion: limiting the Point Source Catalogue}

\begin{figure}[htbp]
\begin{center}
\resizebox{8.5cm}{!}{ \rotatebox{0}{\includegraphics{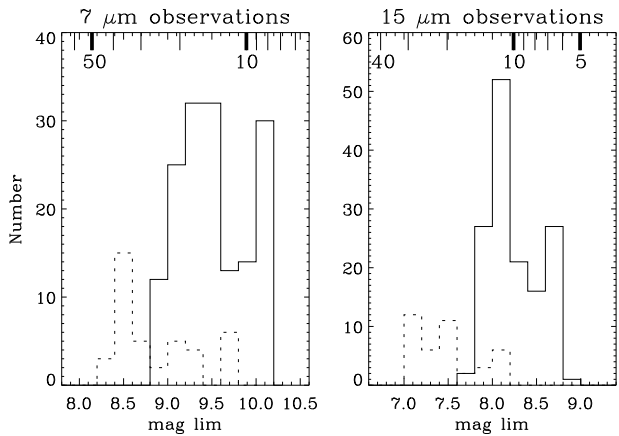}}}
\caption{Distribution of the magnitudes mag$_{\rm lim}$ at which the catalogues
have been cut for the broad filters LW2 and LW3 (full lines), and for
the narrow filters (dotted lines). However, note that, for the narrow filters,
the data with magnitudes higher than mag$_{\rm lim}$-0.5 are of poor quality
(see text, Sect. \ref{artif}~B).
The logarithmic scales at the top of each panel show the corresponding
flux densities in mJy for LW2 and LW3. A small correction has to be
applied for the corresponding flux densities with narrow filters (see
Table~\ref{filters}).}
\label{distrib_mag_lim}
\end{center}
\end{figure}

The distribution of the limiting magnitudes, as defined in the previous section
(Eqs. (\ref{mag_lim6_1}), (\ref{mag_lim6_2}), (\ref{mag_lim6_3})
and (\ref{mag_lim6_4}) for 6'' pixel observations, Table~\ref{tab-mag_lim3}
for 3'' pixel observations) for all ISOGAL observations is shown in
Fig.~\ref{distrib_mag_lim}. Since most observations were done with the
broad LW2 and LW3 filters, these histograms show that the typical reached
sensitivity is around 20~mJy at 7~$\mu$m and 12~mJy at 15~$\mu$m.

When we apply these relations  to all the ISOGAL catalogues, we eliminate
$\approx$25\% of the sources. This photometric cut is far more severe for
moderate quality sources
than for good quality ones: if we consider the QUALITY flag as defined
in Sect.~\ref{psc-mesh}, it appears that about one half of the sources
with QUALITY~=~1 or 2 are discarded, while $\sim$~30\% of the sources
with QUALITY~=~3 and $\sim$~12\% of the sources with QUALITY~=~4 are
removed by this cut.

\subsection{Repeated observations}

\subsubsection{\label{sec-repeated}Overlapping 6'' observations}

A few ISOGAL fields have been observed twice or more with exactly the same
observational setup (filter and pixel size), and a large
number of fields have overlapping regions. The total surface of such
repeatedly observed areas is $\sim$ 0.7 deg$^2$.
A comparison of the photometry extracted from such independent observations
of the same regions of the sky was performed, and the main results for each
observational setup are given in Table~\ref{repeated}. Note that,
because of the variability of some sources,
the quoted standard deviations in Table~\ref{repeated}
are slightly above the true photometric uncertainty of the final
catalogue multiplied by $\sqrt{2}$.

\begin{table}[htbp]
\caption[]{\label{repeated} Main results of the comparison of repeated
observations}
\begin{center}
\begin{tabular}{|l|cccc|}
\hline
Filter&Overlap&Nb. of&$\langle \Delta$mag$\rangle$&RMS \\
 &surface (deg$^2$)&sources& & \\
\hline
LW2 & 0.166 & 2793 & 0.008 & 0.21 \\
LW6 & 0.098 & 1974 & 0.005 & 0.22 \\
\hline
LW3 & 0.275 & 2244 & 0.009 & 0.23 \\
LW9 & 0.111 & 1250 & 0.007 & 0.28 \\
\hline
Total & 0.650 & 8261 & 0.003 & 0.23 \\
\hline
\end{tabular}
\end{center}
\end{table}


It is also possible to get information about the completeness from the
fraction of sources detected in both overlapping observations.
It is however difficult to accurately estimate the completeness
level by this method, as neither of the two catalogues is complete.
It is nevertheless possible to have a rough estimate by comparing the catalogue
extracted from a 6'' pixel observation with the more complete one, extracted
from a 3'' pixel observation of the same region. Then we can compute the
magnitude above which the fraction of 3'' sources found in the 6'' catalogue
is below 50\%. Taking into account all the limitations inherent to this
method, the final results are essentially consistent with those derived from
the artificial sources simulations, and also confirm that more care
has to be taken for the observations performed with narrow filters.

\subsubsection{\label{sec-reality}Reality of the extracted sources}

An additional check of the reality of the sources can be performed as
follows. The sources extracted from 6'' pixel observations should also
be found in a 3'' pixel observation of the same region, because the
sensitivity is generally greater in the latter, since the source extraction
is much less limited by confusion. Also sources detected
at one wavelength and with a good quality association at another
ISO or DENIS wavelength have a very large probability to be real.
But sources found only in a 6'' pixel observation, with counterparts
neither in the overlapping 3'' pixel observation nor at other wavelengths
(or with a bad quality association) may be spurious.

From the available set of overlapping 3'' and 6'' observations,
we have determined that the overall
fraction of such doubtful sources is very small ($\sim$7\%),
with a large difference between the 7~$\mu$m ($\sim$4\%) and
the 15~
$\mu$m ($\sim$11\%) sources. This fraction also strongly
depends on the quality of the sources, and ranges from less than
1\% (at both wavelengths) for sources with quality flags Q~=~4,
to $\sim$15\% (resp. $\sim$30\%) for sources with Q~=~1 or
2 or with MESH~=~1 or 2 at 7~$\mu$m (resp. at 15~$\mu$m).
Therefore sources with quality flags less than 3 should be
considered with extreme caution, especially at 15~$\mu$m.

\subsection{\label{sec-715-assoc}7-15~$\mu$m cross-identification}

\subsubsection{\label{sec-7-15-astrom}Astrometric correction}

The initial astrometric accuracy of the ISOCAM data is limited by the errors
in the pointing of the telescope and in the positioning of the lens wheels.
The global astrometric uncertainty can reach $\sim$~10'' (Blommaert et al.
\cite{isocam_hb}, see also Ott \cite{Ott}), and the
offset between two independent observations can reach twice this value.
Therefore an offset correction between the 7~$\mu$m and the 15~$\mu$m
observations was needed before the two catalogues could be cross identified.
The found offsets are typically of order a few arcseconds, but can
reach 15'', in agreement with expectations.

In addition, there can be a small error in the positioning of the individual
images within the final raster, due to a combination of possible long term
drifts and the lens wheel jitter. Only very small amplitude ``distortion''
effects have been observed, but a low order polynomial correction was
systematically applied to the 15~$\mu$m coordinates to best match the
7~$\mu$m ones.

\subsubsection{Source associations}

\begin{figure}[htbp]
\begin{center}
\resizebox{8.8cm}{!}{ \rotatebox{0}{\includegraphics{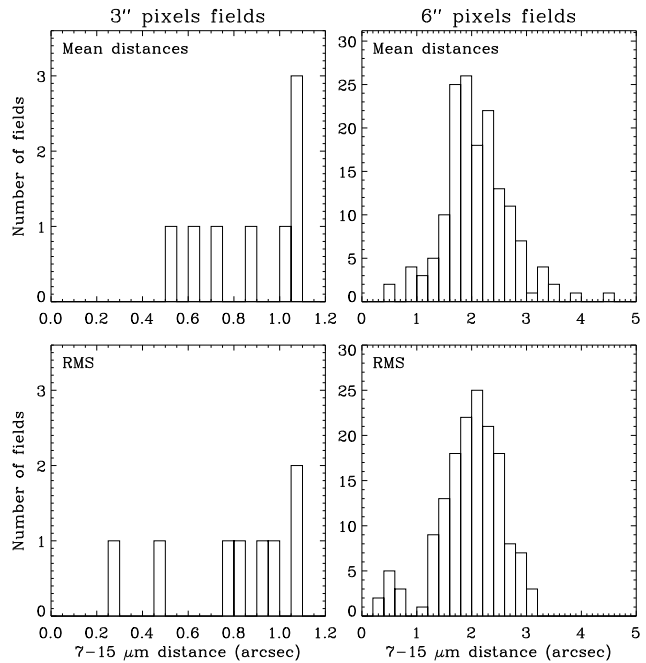}}}
\caption{\label{hist_rms715}
Top panel: distribution of the mean values of the separations between associated
7~$\mu$m and 15~$\mu$m sources after astrometric correction in all ISOGAL FC fields.
Bottom panel: distribution of the standard deviations of these
separations.}
\end{center}
\end{figure}

After the 15~$\mu$m coordinates were corrected to match those at
7~$\mu$m, an association between 7~$\mu$m and 15~$\mu$m sources
was performed with a search radius equal to two pixels. This rather large
radius was chosen in order not to miss 7--15~$\mu$m associations for slightly
extended sources, and because the density of 15~$\mu$m sources is low enough
to limit the probability of chance associations to a few percent in most
cases. Only associations with the smallest separation are retained. The
mean values of the 7--15~$\mu$m separations are typically in the range 1--3''
in all ISOGAL FC fields, with standard deviations in the same range, as shown
in Fig.~\ref{hist_rms715}. At the end of this step, the catalogued source
coordinates are the most accurate available, namely the 7~$\mu$m coordinates
for the sources detected at 7~$\mu$m, or the 15~$\mu$m coordinates translated
to the 7~$\mu$m referential for the 15~$\mu$m sources with no 7~$\mu$m
association in the FC fields. We kept the initial 15~$\mu$m coordinates
only for the sources in FB fields without 7~$\mu$m observations.

\subsubsection{\label{sec-flag715}The 7--15~$\mu$m association quality flag}

Finally, a 7--15~$\mu$m association quality flag is computed for each associated
source. The value of this flag is defined as follows:
\begin{description}
\item[4]: the separation between the 7~$\mu$m and the 15~$\mu$m sources is
$\leq$ 1 pixel and there is only one possible association within a radius
of 2 pixels
\item[3]: the separation is still $\leq$ 1 pixel but there is another 15~$\mu$m
source at less than 2 pixels
\item[2]: the separation is between 1 and 2 pixels, and there is no other
source within a radius of 2 pixels
\item[1]: the separation is between 1 and 2 pixels, and there is at least 2
sources within a radius of 2 pixels
\end{description}

\begin{figure}[htbp]
\begin{center}
\resizebox{8.8cm}{!}{ \rotatebox{0}{\includegraphics{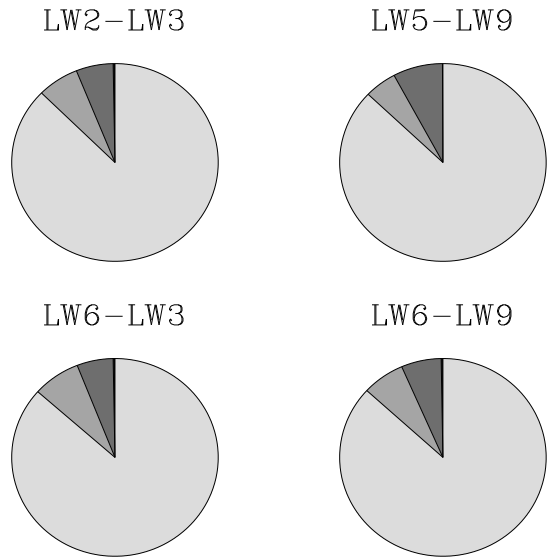}}}
\caption{\label{pie_flag715}
Distribution of the values of the 7--15~$\mu$m association quality flag for the
different combinations of 7 and 15~$\mu$m filters. The gray scale corresponds
to the different values of this flag, from 4 (lightest gray) to 1 (darkest gray).
Only very few sources have this flag equal to 1, so that the darkest gray is
hardly visible in these plots.}
\end{center}
\end{figure}

The distribution of the values of this flag is shown in Fig.~\ref{pie_flag715}.
A very large majority of the associated sources have a very good quality of
association: 87\% of the associations have $Q_{7-15}$=4 and 6.4\% have
$Q_{7-15}$=3.  Only $\sim$~6\% of these flags are equal to 2 and fewer than
0.3\% are equal to 1, corresponding to an association distance larger
than one pixel. However, 19\% of the
sources detected at 15~$\mu$m within the area also observed at 7~$\mu$m have
no association, while 47\% of the 7~$\mu$m sources in the common area have
no 15~$\mu$m counterpart. This large difference is explained by the deeper
sensitivity of the 7~$\mu$m observations, as compared to the 15~$\mu$m ones.

\section{\label{sec_denis}DENIS observations of the central Galaxy}

In addition to these mid-infrared wavelengths, all the observations in
the southern hemisphere (almost 95\% of the total area) have been
systematically cross-identified with the DENIS (Epchtein et al. \cite{DENIS},
\cite{ref-DENIS2})
data, which provide measurements in the three near infrared bands $I$, $J$
and $K_{\rm s}$.

\subsection{The DENIS ``Bulge'' project (Simon et al., in preparation)}

In coordination with the ISOGAL project, dedicated observations with the DENIS
instrument on the ESO 1 meter telescope at La Silla have been performed, along
the inner Galactic plane, between -30 and +10 degrees in galactic longitude, -2
and +2 degrees in latitude, ($\pm$ 4 degrees in the inner Bulge) using a
specific technique (Simon et al. in preparation). The individual images
(12'x12') were taken in a raster mode, covering typically 3 square degrees.
Between +10 and +30 degrees in longitude, regular
30$^{\circ}$ DENIS strips (see Epchtein et al. \cite{DENIS}) were used, with a
special reduction procedure.
All the DENIS images which have been used to build the ISOGAL PSC are
described in the Table of DENIS Observations, whose format is given in
Table~\ref{denis_tab}.

\begin{table*}[htbp]\normalsize
\caption{\label{denis_tab}Format of DENIS observations (12' $\times$ 12' images)
Table (version 1)}
\begin{center}
\begin{tabular}{|rllll|} 
\hline
col & name & format & units [range] & description \\
\hline
 1 & Name      & a7   &                  & image number \\
 2 & date      & a6   & YYMMDD           & date of observation \\
 3 & j\_day    & i4   &                  & Julian day of observation - 2450000 \\
 4 & RA        & f8.4 & deg              & RA (J2000) of image centre \\
 5 & Dec       & f8.4 & deg              & Dec (J2000) of image centre \\
 6 & G\_lon    & f7.3 & deg [-180--+180] & Galactic longitude of image centre \\
 7 & G\_lat    & f7.3 & deg [-90--+90]   & Galactic latitude of image centre \\
 8 & q\_I      & i1   &                  & quality flag of $I$ image \\
 9 & q\_J      & i1   &                  & quality flag of $J$ image \\
10 & q\_K      & i1   &                  & quality flag of $K_{\rm s}$ image \\
\hline
\end{tabular}
\end{center}
\end{table*}

\subsection{\label{denis_astrom}Data processing and accuracy}

The source extraction has been made through PSF fitting, using the same
extraction code as for ISOCAM images. The PSF is modelled in 9 squares
on each 12'$\times$12' individual image and adjusted with
respect to the source position. The derived correlation factor gives an
evaluation of the photometric uncertainty of the source extraction. For each
band, we preserve only the sources with a correlation factor greater than 0.6.
The correlation factors are given for each DENIS source in the ISOGAL PSC
(Sect.~\ref{sec_psc}).

The saturation of DENIS detectors occurs around magnitude 10 in $I$, 7.5 in
$J$ and 6 in $K_{\rm s}$, and results in severely underestimated flux densities.
Therefore, the brightest DENIS sources have been removed from the catalogue.
The absolute photometry results from the zero point derived from standard stars
observed
through the night. A mean value is applied. These magnitudes can be converted
to flux densities using the zero points given in Table~\ref{zp_denis}
(from Fouqu\'e et al. \cite{Fouque}).

\begin{table}
\begin{center}
\caption{\label{zp_denis}Isophotal wavelengths and zero point flux densities
for the three DENIS bands}
\begin{tabular}{|ccc|}
\hline
Band & $\lambda_{\rm iso}$ ($\mu$m) & F$_{\nu}$ (Jy) \\
\hline
$I$ & 0.791 & 2499 \\
$J$ & 1.228 & 1595 \\
$K_{
m s}$ & 2.145 & 665 \\
\hline
\end{tabular}
\end{center}
\end{table}

The limiting sensitivity is about 0.05~mJy (mag. 19) in $I$, 0.5~mJy (mag. 16)
in $J$ and 2.5~mJy (mag. 13.5) in $K_{\rm s}$ but the extraction can become
confusion limited in the dense Galactic environment. The relative accuracy of
the photometry is checked through the comparison of the measurements in the
overlaps (2' between adjacent images).
The average differences are better than 0.03 mag down to magnitudes 17 in $I$
(standard deviation $<$0.1~mag), 14 in $J$ and 12 in $K_{\rm s}$ (standard deviation
$<$0.2~mag), which remains very good given the difficulty inherent to such
dense regions.

Finally, an image quality flag has been evaluated from the overlapping
regions of each DENIS frame covering the ISOGAL rasters. In each band the
standard deviation of magnitude differences over a defined magnitude range
is calculated (see Table~\ref{def_denis_flag}) and we assigned a quality
flag ranging from 0 (very bad) to 3 (very good). More than one half of the
used images have this flag equal to 4, and less than 20\% have a flag
equal to 1 or 0.

\begin{table}
\begin{center}
\caption{\label{def_denis_flag}Definition of the DENIS image quality flags}
\begin{tabular}{|cccccc|}
\hline
\multicolumn{6}{|c|}{ \hspace{-1.5cm} mag. range \hspace{2.0cm} sigma range}\\
Flag &     &0&1&2&3\\
\hline
$I$ & 11-16 & $>0.15$ & 0.1 -0.15 & 0.07-0.1  & $\rm < 0.07$ \\
$J$ & 9-14 & $>0.20$ & 0.16-0.20 & 0.13-0.16 & $\rm < 0.13$ \\
$K_{\rm s}$ & 7-12 & $>0.20$ & 0.16-0.20 & 0.13-0.16 & $\rm < 0.13$ \\  
\hline
\end{tabular}
\end{center}
\end{table}


The astrometry is calculated for each image from the present
association between $I$ and
the USNO\_A2 catalogue. Then, the cross associations of $J$ data over $I$, and
of $K_{\rm s}$ data over $J$ are relatively straightforward since all three images
have been observed simultaneously. The resulting relative accuracy is better
than 0.2''
(RMS) in $I$ and 0.4'' in $J$ and $K_{\rm s}$. The derived position for $I$ is kept
for $I$/$J$/$K_{\rm s}$ associations, and the $J$ position is given for the
$J$/$K_{\rm s}$
associated sources. From a comparison made with the TYCHO catalogue in the SgrI
field in the Baade's Window, no systematic offset was found. The mean value of
the distances was 0.36'', with a 0.19'' standard deviation (Simon et al.,
in preparation).
Altogether the present accuracy of the DENIS coordinates used is thus
better than 0.5''. It will be improved in the future since it is greatly
limited by the accuracy of the astrometry of the USNO\_A2 catalogue.

\subsection{\label{assoc_denis}ISOGAL--DENIS cross-identification}

The general method that we used to associate DENIS sources with ISOGAL
sources is similar to the procedure we used to associate 7~$\mu$m and 15~$\mu$m
data. The only difference arises from the very high density of DENIS sources,
so that we used a much smaller association radius, and we cut out the faintest
DENIS sources when the source density was too high, in order to reduce the
probability of chance associations.

\subsubsection{Astrometric correction}

As explained in Sect.~\ref{denis_astrom}, the absolute accuracy of the
DENIS coordinates is better than 0.5'', thus much better than the
ISO astrometry. Therefore we took the DENIS coordinates as the
reference system, and computed the global translation offset between the
ISOGAL and the DENIS catalogues with the same procedure as for
the 7--15~$\mu$m associations. The resulting offsets are typically in the
range 3--9'', and can be explained
by the lens wheel jitter of ISOCAM (Sect.~\ref{sec-7-15-astrom}). This
also implies that the coordinates of ISOGAL sources outside the region with
DENIS observations can be wrong by this range of distances.
An approximate polynomial distortion correction was computed with the same
procedure as for the 7--15~$\mu$m associations, in order to match as best as
possible the previous ISO reference coordinates with the DENIS ones. Again,
the observed effects were of very small amplitude, but this correction was
required to correct for small rotations in the ISOCAM rasters.

\subsubsection{Confusion cut of weak DENIS sources}

The catalogue of DENIS sources covering each ISOGAL field was first
limited to sources
with a $K_{\rm s}$ detection, since a $J$--7~$\mu$m association without
$K_{\rm s}$ counterpart has a large probability of being a misidentification.
The density remains very high at this stage,
exceeding 10$^5$ sources/deg$^2$
in the Galactic Centre region. Therefore we further cut the DENIS catalogue
to a $K_{\rm s}$ magnitude that gave a source density of 72~000 sources/deg$^2$
for the ISO 3'' pixel observations. For the observations with 6'' pixels,
we proceeded in two steps, first limiting the DENIS source density to
18~000 sources/deg$^2$ and then to 36~000 sources/deg$^2$ (see below).
This confusion cut, with the procedure described
below, enabled us to limit the probability of
chance associations to a few percent even in the most crowded fields.

\subsubsection{Source associations}

The search for DENIS associations was done with the same procedure as for the
7--15~$\mu$m associations, with a smaller search radius.
The mean values of ISO--DENIS separations that we found are
typically in the 1--2'' range for all ISOGAL fields, with a
few larger values for the FB fields, in which the association is done between
DENIS and 15~$\mu$m coordinates (see e.g. Fig. E-31 in the electronic version
of this paper). The corresponding standard deviations are
mainly in the 1--1.5'' range. An association radius of $\sim$ 3--4'' is thus
appropriate to find most good associations with a low probability of
spurious results. However, a close inspection of the distribution
of association radii shows that, in a few fields with poor data quality, a few
real associations may have a larger association radius, in particular for
blended or extended sources with 6'' pixels. Therefore, for the ISO 3''/pixel
observations, we used a 3.6'' search radius. But for the ISO 6''/pixel
observations, we pushed the search up to a radius of 7''; however, we carefully
distinguished by quality flags the associations with separations smaller or
larger than 3.5''.

With such values, the probabilities of random associations may appear
high. However, as discussed below, because of the large fraction of real
associations with smaller separation, the actual fraction of spurious
associations with reasonably good quality flags remains lower than a few
percent. The chance of spurious association is larger for weaker $K_{\rm s}$
sources allowed with the higher density limit. The final ISO--DENIS quality
flag (Sect.~\ref{sec-flagID}) takes this point into account.

\subsubsection{\label{sec-flagID}The ISO--DENIS association flag}

The ISO--DENIS association is characterised by a specific quality flag,
$Q_{\rm ID}$, which ranges in values from 5 (highest quality) to 0
(no association). The computation of
this flag takes into account:
\begin{itemize}
\item
the separation between the ISO source and the associated closest
DENIS source
\item
the number of DENIS sources within the search radius
\item
the global quality of ISO--DENIS associations for each field, as derived from
a visual inspection of the histograms of the distances of associations
\item
for the ISO 6'' observations only, the value of this flag is decreased for
sources with a $K_{\rm s}$ magnitude between the two cutoff magnitudes K\_max1 and
K\_max2 (Cols. 16 and 17 of the table of ISOGAL fields, see
Table~\ref{fields-cat}), which were used to limit the source density
of the DENIS catalogue to 18~000 and 36~000 sources per square degree,
respectively.
\end{itemize}

\begin{figure}[htbp]
\begin{center}
\resizebox{8.8cm}{!}{ \rotatebox{0}{\includegraphics{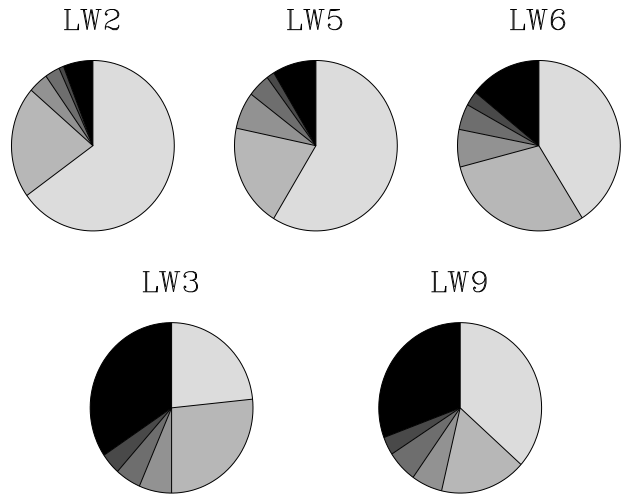}}}
\caption{\label{pie_flagID}
Distributions of the ISO--DENIS association flag for the
different ISO filters. The gray scale corresponds
to the different values of this flag, from 5 (lightest gray) to 1 (darkest gray),
and the black sectors show the fraction of sources without DENIS association
within the area observed by DENIS.}
\end{center}
\end{figure}

Let us stress the large fraction of DENIS associations, 
$\sim$92\% for 7~$\mu$m sources, $\sim$79\% for 15~$\mu$m sources in
FB fields and $\sim$45\% for 15~$\mu$m sources with no 7~$\mu$m association
in FC fields (Fig. \ref{pie_flagID}). The fraction of associations with
K\_max1 $< K_s <$ K\_max2 is also small, $\sim$4\% for 7~$\mu$m sources,
$\sim$2.5\% for 15~$\mu$m sources in FB fields and $\sim$17\% for 15~$\mu$m
sources with no 7~$\mu$m association in FC fields.
Therefore, the fraction of spurious associations among accepted
associations (see below) always remains small, typically at most
$\sim$1\% for 7~$\mu$m sources and a few percent for 15~$\mu$m sources.

Finally when the derivation leads to $Q_{\rm ID}\,=\,$ 0, the association is
considered as invalid and no DENIS association is
given in the catalogue. With this
definition, associations with a quality flag equal to 4 or 5 can be considered
as secure, while a value of 3 is more uncertain but remains a high probability
association, and values of 1 or 2 are more doubtful but still include an
appreciable fraction of real associations. The distribution of
the computed ISO--DENIS association flags is shown in Fig.~\ref{pie_flagID},
where it can be seen that $\sim 87\%$ of the associations found have a good
quality (flag $\geq$ 4), while fewer than 8\% of the 7~$\mu$m sources (LW2,
LW5 and LW6 filters) within the area observed by DENIS have no association.

\section{\label{sec_psc}ISOGAL--DENIS Point Source Catalogue (version 1)}

The Point Source Catalogue contains a total of 106~150 sources, and is
composed of two sections. For each field, the ``regular'' catalogue contains
all the sources inside the formal limits of the rectangular field, as defined
in Table~\ref{fields-cat} (see example in Fig. \ref{field_fafc}).
These limits have been
computed to avoid any border effects: all the sources inside this area
are located at more than two pixels from the saw-tooth edges of the observed
raster, both at 7 and 15~$\mu$m for FC fields. This differs from the EDGE flag
computed for each wavelength (see Sect.~\ref{psc-mesh}) since the ``regular''
region is limited to a rectangular area (whose axis are aligned along the
galactic ones) which has been fully observed at both wavelengths.

Then, the ``edge'' catalogue contains the sources outside the limits of the
rectangular field, but excluding the measurements at less than two pixels
from the saw-tooth edges. This means that in the ``edge'' region
of an FC field, it is possible to find a source with for example a 7~$\mu$m
detection and no 15~$\mu$m counterpart, simply because the edges of the
15~$\mu$m raster do not exactly match the ones of the 7~$\mu$m raster, so
that the source can be outside the region observed at 15~$\mu$m or within
2 pixels of one saw-tooth edge.
As a result, $\sim$53\% of the 7~$\mu$m sources and $\sim$81\% of the
15~$\mu$m sources in the ``regular'' regions of all FC fields have
an association at the other ISO wavelength, while these fractions become
$\sim$47\% for 7~$\mu$m sources and $\sim$70\% for 15~$\mu$m sources
in the ``edge'' regions.

Both the ``regular'' and the ``edge'' catalogues have the format described
in Table~\ref{psc}, and a few examples of entries are given in
Table~\ref{psc_example}. The final Catalogue contains 93~385 sources in the
``regular'' regions, and 12~765 sources in the ``edge'' regions.

\begin{table*}[htbp]\normalsize
\begin{minipage}{\textwidth}
\renewcommand{\thefootnote}{\textit{\alph{footnote}}}
\caption{\label{psc}Format of the ISOGAL Point Source Catalogue
(version 1) - 106~150 entries (see examples in Table~\ref{psc_example})} 
\begin{center}
\begin{tabular}{|rllll|} 
\hline
col & name & format & units [range] & description \\
\hline
 1 & {\bf Number } & a5   &                             & source identification number in the field \\
 2 & {\bf Name }   & a25  & ISOGAL-PJhhmmss.s$\pm$ddssmmX & source identifier (J2000)\footnote
{The last character `X' is only present when two sources with the same position have
to be distinguished (see text, Sect. \ref{sec-psc-1})} \\
 3 & {\bf RAJ2000} & f8.4 & deg [0--360]                & Right Ascension (J2000)\footnote
{Coordinates: the final adopted coordinates (Cols. 3 and 4) are the DENIS
ones if there is an association, or the ISO corrected to DENIS if an
observation exists but no source was associated. In the northern fields
(without DENIS), the coordinates are the 7~$\mu$m ones if they exist, or
the 15~$\mu$m ones for the sources in FB fields, and the 15~$\mu$m
corrected to 7~$\mu$m for the sources detected only at 15~$\mu$m in the FC
fields. When no DENIS association exists, RAJ2000$=$RAISOGAL and DEJ2000$=$DEISOGAL.} \\
 4 & {\bf DEJ2000} & f8.4 & deg [-90--+90]              & Declination (J2000) \\
 5 & RAISOGAL      & f8.4 & deg [0--360]                & ISOGAL RA (J2000) \\
 6 & DEISOGAL      & f8.4 & deg [-90--+90]              & ISOGAL Dec (J2000) \\
 7 & G\_lon        & f8.4 & deg [-180--+180]            & Galactic longitude \\
 8 & G\_lat        & f8.4 & deg [-90--+90]              & Galactic latitude \\
 9 & I\_field      & a14  & Fxslllllsbbbbb              & ISOGAL field name \\
10 & D\_field      & a7   &                             & DENIS image
name\footnote{Only the seven last digits of the DENIS numbers have been stored,
as the three first ones are always 000.} \\
\hline 
11 & {\bf Imag }   & f5.2 & mag                         & DENIS $I$-band
magnitude\footnote{A value of 88.88 for a magnitude means that this position
was not observed at this wavelength, while a value of 99.99 means that the source
was not detected at this wavelength.} \\
12 & {\bf Icorr }  & f4.2 & [0--1]                      & DENIS $I$-band correlation factor \\
13 & x\_I          & f5.1 & pixel                       & x-position in DENIS $I$-band image \\
14 & y\_I          & f5.1 & pixel                       & y-position in DENIS $I$-band image \\
15 & {\bf Jmag }   & f5.2 & mag                         & DENIS $J$-band
magnitude\footnotemark[4]{} \\
16 & {\bf Jcorr }  & f4.2 & [0--1]                      & DENIS $J$-band correlation factor \\
17 & x\_J          & f5.1 & pixel                       & x-position in DENIS $J$-band image \\
18 & y\_J          & f5.1 & pixel                       & y-position in DENIS $J$-band image \\
19 & {\bf Kmag }   & f5.2 & mag                         & DENIS $K_{\rm s}$-band
magnitude\footnotemark[4]{} \\
20 & {\bf Kcorr }  & f4.2 & [0--1]                      & DENIS $K_{\rm s}$-band correlation factor \\
21 & x\_K          & f5.1 & pixel                       & x-position in DENIS $K_{\rm s}$-band image \\
22 & y\_K          & f5.1 & pixel                       & y-position in DENIS $K_{\rm s}$-band image \\
\hline
23 & {\bf mag7 }   & f5.2 & mag                         & ISOGAL 7~$\mu$m
magnitude\footnotemark[4]{} \\
24 & {\bf e\_mag7} & f4.2 & mag                         & uncertainty in 7~$\mu$m magnitude \\
25 & filt\_7       & i1   & [2,5,6]                     & LW number of filter used \\
26 & pfov\_7       & i1   & arcsec [3,6]                & pixel field of view \\
27 & x\_7          & f6.2 & pixel                       & x-position on ISOGAL final 7~$\mu$m image \\
28 & y\_7          & f6.2 & pixel                       & y-position on ISOGAL final 7~$\mu$m image \\
29 & npix\_7       & i1   & [0--7]                      & npix flag at 7~$\mu$m (see Sect.~\ref{psc-mesh}) \\
30 & mesh\_7       & i1   & [1,2,3]                     & mesh flag at 7~$\mu$m (see Sect.~\ref{psc-mesh}) \\
31 & edge\_7       & i1   & [0,1]                       & edge flag at 7~$\mu$m (see Sect.~\ref{psc-mesh}) \\
32 & {\bf qual\_7} & i1   & [0--4]                      & global quality flag at 7~$\mu$m
(see Sect.~\ref{psc-mesh}) \\
33 & {\bf mag15 }  & f5.2 & mag                         & ISOGAL 15~$\mu$m
magnitude\footnotemark[4]{} \\
34 & {\bf e\_mag15}& f4.2 & mag                         & uncertainty in 15~$\mu$m magnitude \\
35 & filt\_15      & i1   & [3,9]                       & LW number of filter used \\
36 & pfov\_15      & i1   & arcsec [3,6]                & pixel field of view \\
37 & x\_15         & f6.2 & pixel                       & x-position on ISOGAL final 15~$\mu$m image \\
38 & y\_15         & f6.2 & pixel                       & y-position on ISOGAL final 15~$\mu$m image \\
39 & npix\_15      & i1   & [0--7]                      & npix flag at 15~$\mu$m (see
Sect.~\ref{psc-mesh}) \\
40 & mesh\_15      & i1   & [1,2,3]                     & mesh flag at 15~$\mu$m (see
Sect.~\ref{psc-mesh}) \\
41 & edge\_15      & i1   & [0,1]                       & edge flag at 15~$\mu$m (see
Sect.~\ref{psc-mesh}) \\
42 & {\bf qual\_15}& i1   & [0--4]                      & global quality flag at 15~$\mu$m
(see Sect.~\ref{psc-mesh}) \\
\hline
43 & dis\_II       & f5.2 & arcsec                      & separation 7 to 15~$\mu$m associated sources \\
44 & {\bf ass\_II} & i1   & [0--4]                      & 7--15~$\mu$m association quality flag \\
45 & dis\_ID       & f5.2 & arcsec                      & separation ISOGAL to DENIS associated sources \\
46 & {\bf ass\_ID} & i1   & [0--5]                      & ISOGAL--DENIS association quality flag \\
\hline
\end{tabular}
\end{center}
\end{minipage}
\end{table*}

\subsection{\label{sec-psc-1}Position data}
The first ten entries for each source in the PSC consist of
general data, as described below.
\begin{itemize}
\item
Col. 1: source number in the field. This number increases with the right
ascension of the sources. Each individual catalogue (the ``regular''
and the ``edge'' for each field) contains its own numbering, and these
numbers are preceded by an ``E'' in the ``edge'' catalogues.
\item
Col. 2: source name. It is composed of 25 characters, following
the format:
\[ ISOGAL-PJhhmmss.s \pm ddssmmX \]
where ``ISOGAL'' stands
for the ISOGAL--DENIS data, the ``P'' means that these are
provisory data, and the Jhhmmss.s$\pm$ddssmm are the J2000
equatorial coordinates of the source, as they appear in Cols.
3 and 4. The last character, `X', is left blank in all cases but those where
two (or exceptionally three) sources from different fields are found at
the same position, because they are associated with the same DENIS source
and because of edge effects. This concerns 842 sources (0.8\% of the PSC)
and in all those cases, at least one of the coinciding sources is in an
``edge'' catalogue. A letter is appended to the name of the sources,
starting with an $a$ for that in a ``regular'' catalogue if it exists,
otherwise using an arbitrary order between the ``edge'' catalogues, and
going to $b$ or $c$ when needed.
\item
Cols. 3 and 4: reference J2000 equatorial coordinates,
expressed in decimal degrees (see the footnote $b$ in Table~\ref{psc}).
\item
Cols. 5 and 6 give the ISOGAL corrected coordinates, which are the
ISOGAL extracted coordinates when there is no DENIS observation of
the field, or the ISOGAL corrected to DENIS system ones when
a DENIS observation exists.
\item
Cols. 7 and 8 give the galactic reference coordinates corresponding
to the reference coordinates given in Cols. 3 and 4, in the commonly
used $(l^{II},b^{II})$ galactic system.
\item
Col. 9 gives the name of the ISOGAL field.
\item
Col. 10 gives the last seven digits of the number of the DENIS image
where an ISO--DENIS association was found. For ISOGAL sources with no
DENIS counterpart, this column contains 0000000.
\end{itemize}

\subsection{DENIS data}
All the DENIS data are given in Cols. 11 to 22. For each of the
three bands, these data are the measured magnitude, the correlation
factor with the PSF, and the pixel coordinates of the source in the
individual DENIS 12'$\times$12' image, whose reference number is given
in Col. 10.

For the ISOGAL sources within the area observed by DENIS but with no
DENIS association, the $I$, $J$ and $K_{\rm s}$ magnitudes are set to 99.99, 
while they are set to 88.88 for all the sources located outside
the region surveyed by DENIS. In these two cases,
the PSF correlation factors and pixel coordinates are set to 0.

The correlation factors with the PSF give an indication of the
photometric quality (see Simon et al., in preparation):
the uncertainty on the measured magnitude
is small when this factor is $\geq$~0.95. On the other hand,
a value $\leq$~0.85 means that the photometry is more uncertain
(typically by 0.1 to 0.2~magnitude). For bright sources, this
may come from moderate saturation effects, while for faint sources,
a value $\leq$0.80 is more typical. Nevertheless, a factor
$\leq$0.70 indicates a poor photometric quality, which
may be caused by blending effects or confusion with the
background.

\subsection{ISOCAM data}
Cols. 23 to 42 give all the data derived from individual 7 and 15~$\mu$m
ISO observations, including quality flags (see Sect.~\ref{psc-mesh}),
calibrated magnitudes, uncertainties ($\sigma$) from the PSF fit measurement
of the magnitudes,
pixel positions in the final image (after correction of the orientation,
see Sect.~\ref{new-images}), filter numbers and pixel sizes.

\subsection{\label{qual_flag}Association quality flags}

\begin{table*}[htbp]\normalsize
\caption{\label{psc_example}Examples of entries in the ISOGAL--DENIS
Point Source Catalogue from the C32 field} 
\begin{center}
\begin{tabular}{|rl|ccc|} 
\hline
col & name & Example 1 & Example 2 & Example 3 \\
\hline
 1 & {\bf Number } & 0008 & 0017 & 0007 \\
 2 & {\bf Name }   & ISOGAL-PJ174118.0-282916 & ISOGAL-PJ174122.7-283146 &
ISOGAL-PJ174117.6-282901 \\
 3 & {\bf RAJ2000} & 265.3250 & 265.3446 & 265.3234 \\
 4 & {\bf DEJ2000} & -28.4880 & -28.5296 & -28.4838 \\
 5 & RAISOGAL      & 265.3251 & 265.3446 & 265.3230 \\
 6 & DEISOGAL      & -28.4880 & -28.5296 & -28.4837 \\
 7 & G\_lon        &  -0.1158 &  -0.1419 &  -0.1129 \\
 8 & G\_lat        &   1.0415 &   1.0048 &   1.0449 \\
 9 & I\_field      & FC+00000+00100 & FC+00000+00100 & FC+00000+00100 \\
10 & D\_field      & 0955338 & 0000000 & 0955339 \\
\hline 
11 & {\bf Imag }   & 16.49 & 99.99 & 16.36 \\
12 & {\bf Icorr }  & 0.96 & 0.0 & 0.91 \\
13 & x\_I          & 367.7 & 0.0 & 376.6 \\
14 & y\_I          & 153.8 & 0.0 & 735.6 \\
15 & {\bf Jmag }   & 10.87 & 99.99 & 10.63 \\
16 & {\bf Jcorr }  & 0.99 & 0.0 & 0.98 \\
17 & x\_J          & 369.9 & 0.0 & 371.1 \\
18 & y\_J          & 154.9 & 0.0 & 745.1 \\
19 & {\bf Kmag }   & 8.32 & 99.99 & 8.02 \\
20 & {\bf Kcorr }  & 0.98 & 0.0 & 0.99 \\
21 & x\_K          & 368.1 & 0.0 & 370.5 \\
22 & y\_K          & 151.8 & 0.0 & 750.9 \\
\hline
23 & {\bf mag7 }   & 7.60 & 3.47 & 7.36 \\
24 & {\bf e\_mag7} & 0.03 & 0.01 & 0.03 \\
25 & filt\_7       & 2 & 2 & 2 \\
26 & pfov\_7       & 6 & 6 & 6 \\
27 & x\_7          & 165.76 & 181.81 & 164.10 \\
28 & y\_7          & 71.76 & 50.07 & 74.02 \\
29 & npix\_7       & 2 & 0 & 2 \\
30 & mesh\_7       & 3 & 3 & 3 \\
31 & edge\_7       & 0 & 0 & 0 \\
32 & {\bf qual\_7} & 4 & 4 & 4 \\
33 & {\bf mag15 }  & 99.99 & 1.54 & 5.84 \\
34 & {\bf e\_mag15}& 0.00 & 0.03 & 0.06 \\
35 & filt\_15      & 0 & 3 & 3 \\
36 & pfov\_15      & 0 & 6 & 6 \\
37 & x\_15         & 0.00 & 181.64 & 163.77 \\
38 & y\_15         & 0.00 & 49.86 & 73.49 \\
39 & npix\_15      & 0 & 1 & 2 \\
40 & mesh\_15      & 0 & 3 & 3 \\
41 & edge\_15      & 0 & 0 & 0 \\
42 & {\bf qual\_15}& 0 & 4 & 4 \\
\hline
43 & dis\_II       & 0.00 & 1.06 & 1.16 \\
44 & {\bf ass\_II} & 0 & 4 & 4 \\
45 & dis\_ID       & 0.32 & 0.00 & 1.25 \\
46 & {\bf ass\_ID} & 5 & 0 & 5 \\
\hline
\end{tabular}
\end{center}
\end{table*}

The value of the ISOGAL 7--15~$\mu$m association flag (see definition in
Sect.~\ref{sec-flag715}) is given in Col. 44, and the separation (in arcseconds)
between the 7~$\mu$m and the 15~$\mu$m positions (after correction of the field
offset) is given in Col. 43. This flag and the corresponding separation are set to
zero for sources with no 7--15~$\mu$m association.

For the ISO--DENIS association, the quality flag (see definition in
Sect.~\ref{sec-flagID}) is given in Col. 46, and the separation (in arcseconds)
between the ISO and the DENIS positions (after correction of the field offset)
is given in Col. 45. Again, these two entries are set to zero when there is
no ISO--DENIS association.

\subsection{Examples}
Table~\ref{psc_example} shows three examples of entries in the ISOGAL--DENIS
Point Source Catalogue. These sources are located in the ``C32'' field
($l=0.0$, $b=+1.0$). The first
one has been detected at 7~$\mu$m but not at 15~$\mu$m, and has a DENIS
association. The second one has been detected at 7 and 15~$\mu$m but has
no DENIS association. Finally, the third one is detected in all five bands.

\section{\label{spurious}Catalogue of spurious sources}

\begin{table*}[htbp]\normalsize
\caption{\label{tab-spurious}Format of spurious sources Table (version 1)}
\begin{center}
\begin{tabular}{|rllll|} 
\hline
col & name & format & units [range] & description \\
\hline
 1 & Number   & a4   &               & identification number in the ION \\ 
 2 & RAJ2000  & f8.4 & deg [0--360]  & ISOGAL RA (J2000) \\
 3 & DEJ2000  & f8.4 & deg [-90--+90]& ISOGAL Dec (J2000) \\
 4 & Mag      & f5.2 & mag           & ISOGAL magnitude \\
 5 & ION      & a8   &               & ISO observation number \\
 6 & x        & f6.2 & pixel         & x-position on ISOGAL final image \\
 7 & y        & f6.2 & pixel         & y-position on ISOGAL final image \\
\hline
\end{tabular}
\end{center}
\end{table*}

As explained in Sect.~\ref{sec-spurious}, three kinds of extracted
sources brighter than the limiting magnitude of each field
are considered spurious: (1) the sources found only in the
'inversion' processed raster, with no counterpart in a 1 pixel search
radius in the 'vision' raster, (2) the sources with simultaneously a
doubtful inversion-vision association (with a separation between 0.5
and 1 pixel) and with a poor detection confirmation (i.e. with no
association between the $mesh=1$ and the $mesh=2$ results), and (3)
the possible remnants of bright sources, found by a procedure that
looked at the same pixel location in the five
successive images of the implied raster.

These sources are published in three distinct tables. Their format is
defined in Table~\ref{tab-spurious}. The numbers, as they appear
in Col. 1, are preceded by an ``I'' for the ``inversion-only''
sources, by an ``M'' for the sources of the second class and by
an ``R'' for the probable remnants.

Note that most spurious sources of
the first two kinds are probably artifacts, but can also be related to
faint extended structures, for which different parameters in the
extraction process result in slightly different coordinates. The third
class of spurious sources is essentially composed of spurious remnants,
but may contain a few real sources, which have been accidentally discarded
by the procedure because of spatial coincidence with a putative remnant.

\section{\label{new-images}ISOCAM corrected images}

The ISOCAM images have been initially processed using version 7.0 of
the off-line processing pipeline (Sect.~\ref{sec-isocam}).
Similar images processed with the latest version of OLP
are now publically available through the Data Archive on the
ISO web site\footnote{http://www.iso.vilspa.esa.es/ida/index.html}.
However, we make available here the OLP7 images together with version 1
of the PSC for consistency, because they have been used for the
extraction of the sources of this catalogue. Improved ISOGAL images
(Miville-Desch\^enes et al. \cite{ref-mamd} and in preparation) will
be published with version 2 of the catalogue.

Because of the difference in orientation between the individual images
(aligned along the satellite axis, thus with the equatorial coordinates)
and the mosaiced rasters (aligned along the galactic axes), and of different
times of observations, the orientation obtained
after the OLP7 processing was different from one raster
to another. We therefore decided to change this orientation
if necessary, in order to use the same convention for
all rasters, and set the orientation to $l$ along decreasing X
and $b$ along increasing Y.

A more important improvement provided by the construction of the ISOGAL PSC
deals with the astrometry, which has been tied to DENIS whenever possible.
The offsets
that we applied to the source coordinates in order to associate the
ISO sources with DENIS have also been applied to the rasters, as
indicated in Table~\ref{ISO-obs}.
For the FC fields with no DENIS observation, the astrometry of
the 15~$\mu$m rasters has been tied to that of the 7~$\mu$m ones.
The corrected images are available through the CDS and the
IAP\footnote{http://www-isogal.iap.fr/Fields/index\_tdt.html} server.

\section{\label{sec-summary}Summary}

The first version of the ISOGAL--DENIS Point Source Catalogue contains
a total of 106~000 sources, with one or two magnitude measurements
in the mid-infrared (7 and 15~$\mu$m), and up to three magnitude
measurements in the near-infrared ($I$, $J$ and $K_{\rm s}$ bands of the
DENIS survey, see Table~\ref{psc} and Table~\ref{psc_example}).
The data are presented in two similar tables, corresponding
to the ``regular'' and the ``edge'' regions of the observed fields.
The latter contains the sources from the edges of the ISOCAM rasters, where
border effects can occur, which can lead to non-association between the
two ISO bands.

The typical RMS photometric uncertainty is at most $\sim$0.1~mag for the
DENIS bands, and better than 0.15~mag for the ISO bands in most cases,
but it can reach 0.3~mag for the faintest sources in the densest fields.
For the most numerous fields observed with broad filters, the limiting
magnitudes of the published catalogues range between 8.8
and 10.1 at 7~$\mic$ (with a median value equal to 9.46~mag, or
$F_{\nu} \sim$~15~mJy), and between 7.7 and 8.8 at 15~$\mic$
(median 8.16~mag, $F_{\nu} \sim$~11~mJy), depending on the source density.
For the most difficult fields observed with narrow filters, these
limits range between 8.2 and 9.6 mag at 7~$\mic$ and between
7.0 and 8.2 mag at 15~$\mic$.
These limits are conservative and the fainter sources have been rejected
in the present version of the PSC\footnote{The complete catalogues, including
the faint sources rejected, may be obtained by requesting the ISOGAL PI,
omont@iap.fr}.

The current astrometric accuracy of the DENIS data used is better than 0.5'' (RMS).
The final coordinates (as they appear in Cols. 3 and 4 of the catalogue - see
Table~\ref{psc} - in equatorial J2000 system, in Cols. 7 and 8 in
the galactic system, and in the name of the source, Col. 2)
of all ISOGAL sources with a DENIS counterpart are the DENIS ones, and should
also be accurate to 0.5''. The astrometry of the ISOGAL sources with no DENIS
association, but within the fields observed by DENIS, is also tied to the
DENIS coordinates, and should therefore be accurate to $\sim$2'' (RMS).
Finally, ISOGAL sources located
outside the area surveyed by DENIS may suffer from the lens wheel jitter
of ISOCAM, resulting in a maximum $\sim$10'' systematic offset in the
extracted coordinates.

Several flags have been implemented to characterise the reliability of the
sources, the quality of their photometry and of the associations between
the different bands. An indication of the reliability of the mid-infrared
detection is also given by the $mesh$ flag (Col. 30 for 7~$\mu$m and Col. 40
for 15~$\mu$m, see Table~\ref{psc}). A value of 3 indicates a good reliability
level, while a value of 1 or 2 shows that the extraction was not perfectly
confirmed, making the real point-like nature of the source doubtful.

The global quality of the ISO photometry and reliability of each source
is quantified by one
quality flag for each band. These two flags are given in Col. 32 for 7~$\mu$m
and in Col. 42 for 15~$\mu$m, and range from 1 to 4, the highest value
corresponding to the best quality. Thus sources with quality flags equal to 1
or 2 should be considered with caution.

The quality of the association between the two ISO bands is also characterised
by a specific flag, which appears in Col. 44, together with the separation
of the association in Col. 43.
When this flag is equal to 3 or 4, which means that the separation between
the 7~$\mu$m coordinates and the 15~$\mu$m ones is smaller than one pixel,
the validity of the association is almost certain,
while a value of 1 or 2 means that the association has to be carefully
checked, but it may be a real association for slightly extended sources.

Finally, the quality of the ISO--DENIS association is quantified by a
flag given in Col. 46 (and the ISO--DENIS separation appears in Col.
45). Here, values of 4 or 5 correspond to secure associations, while
a value of 3 means that the association was not straightforward, but
it still has a good probability to be real. When this flag is equal
to 1 or 2, the reality of the association has to be checked carefully,
using for instance colour compatibility criteria.

\section{Conclusion}

With the first public version of the ISOGAL--DENIS Point Source Catalogue,
we provide the astronomical community with a catalogue containing about
10$^5$ mid-infrared sources, detected at 7 and/or 15~$\mu$m in the
obscured centre of the Galaxy. The bulk
of them are associated with near-infrared data from the DENIS survey. We
also provide nearly 400 mid-infrared images, with an astrometric
accuracy of $\sim$1'' for most of them.

All the data were reduced using data products of version 7 of the ISO
off-line processing pipeline. Additional specific procedures enabled us
to greatly reduce the number of artifacts and to reduce the photometric
uncertainty to typically 0.15~mag, at the cost of limiting the published
catalogue in the densest observed fields to levels well above the
sensitivity limit of a few mJy.

A second version of the catalogue is already under development, based
on a systematic reprocessing of the raw data using the most up-to-date
specialised procedures (Miville-Desch\^enes \cite{ref-mamd} and
in preparation). This second version will also
contain systematic cross-associations with the near-infrared data of the
2MASS survey, and with the mid-infrared data of the MSX survey.

\begin{acknowledgements}

The ISOCAM data presented in this paper were analysed using `CIA', a joint
development by the ESA Astrophysics Division and the ISOCAM Consortium. The
ISOCAM Consortium is led by the ISOCAM PI, C. Cesarsky.
We thank A. Abergel, H. Aussel, A. Coulais, R. Gastaud, M. P\'erault,
J.L. Starck and many other members
of the ISOCAM team, of the ISO/ESA team at Villafranca and especially of
the CIA team for their help in the ISOGAL data reduction.
We are very grateful to all people who contributed to the ISOGAL data
reducion, including T. August, X. Bertou, E. Copet and M. Unavane.

This publication made use of data products from the Midcourse Space 
Experiment. Processing of the data was funded by the Ballistic 
Missile Defense Organization with additional support from NASA 
Office of Space Science.

This work was carried out in the context of EARA, the European Association
for Research in Astronomy.

S. Ganesh was supported by a fellowship from the Minist\`ere des
Affaires Etrang\`eres, France, and this research was supported by
the Project 1910-1 of Indo-French Center for
the Promotion of Advanced Research (CEFIPRA). SG also acknowledges
the support he received from the French CNRS for participating in the
astronomical school in Les Houches in 1998. M. Schultheis aknowledges
the receipt of an ESA fellowship. B. Aracil and A. Soive were posted
to the ISOGAL Project by the D\'elegation G\'en\'erale de l'Armement,
France.

We are grateful to Dr.~M.~Cohen for his help in the calibration of ISOCAM
data, and to Dr.~S.~Ott and Prof.~I.S.~Glass for their useful comments
and inputs.

The DENIS project is supported, in France by the Institut National des 
Sciences de l'Univers, the Education Ministry and the Centre National de la 
Recherche Scientifique, in Germany by the State of Baden-W\"urtemberg, in 
Spain by the DGICYT, in Italy by the Consiglio Nazionale delle Ricerche, in 
Austria by the Fonds zur F\"orderung der wissenschaftlichen Forschung and the
Bundesministerium f\"ur Wissenschaft und Forschung.
We thank all the members of the DENIS team who allowed obtaining the DENIS
data, and especially J. Borsenberger and S. B\'egon for their processing.

\end{acknowledgements}


\begin{thebibliography}{}

\bibitem[1998]{Abergel}
Abergel, A., Miville-Desch\^enes, M.-A., D\'esert, F.-X., et al. 1998,
``The transient behaviour of the long wavelength channel of ISOCAM'',
http://www.iso.vilspa.esa.es/users/
expl\_lib/CAM/transient\_detector\_ws.ps.gz

\bibitem[1998]{Aussel}
Aussel, H. 1998, August 13, ``ISOCAM LW channel Field of View Distortion'',
http://www.iso.vilspa.esa.es/users/
expl\_lib/CAM/distortion.ps.gz

\bibitem[2002]{ref-artif}
Bellazzini, M., Fusi Pecci, F., Montegriffo, P., et al. 2002, AJ 123, 2541

\bibitem[1998]{Biviano}
Biviano, A., Sauvage, M., Gallais, P., et al. 1998, May 18, ``The ISOCAM Dark
Current Calibration Report'', http://
www.iso.vilspa.esa.es/users/expl\_lib/CAM/darkdoc.ps.gz

\bibitem[1998]{photom_rep}
Blommaert, J. A. D. L. 1998, December 18, ``ISOCAM Photometry Report'',
http://www.iso.vilspa.esa.es/users/
expl\_lib/CAM/photom\_rep\_fn.ps.gz

\bibitem[2000]{Blom2000}
Blommaert, J. A. D. L., Metcalfe, L., Altieri, B., et al. 2000, Experimental
Astronomy 10, 241

\bibitem[2001]{isocam_hb}
Blommaert, J. A. D. L., Siebenmorgen, R., Coulais, A., et al. 2001,
``The ISO Handbook, Volume III: CAM -
The ISO Camera'', http://www.iso.vilspa.esa.es/manuals/
HANDBOOK/III/cam\_hb/

\bibitem[2001]{Bontemps}
Bontemps, S., Andr\'e,  P., Kaas, A. A., et al. 2001, A\&A 372, 173

\bibitem[2000]{GPsurvey}
Burgdorf, M. J., Cohen, M., Price, S. D., et al. 2000, A\&A 360, 111

\bibitem[1996]{ISOCAM}
Cesarsky, C. J., Abergel, A., Agn\`ese, P., et al. 1996, A\&A 315, L32

\bibitem[2000]{Coulais}
Coulais, A., \& Abergel, A. 2000, A\&AS 141, 533

\bibitem[2001]{ref-firback}
Dole, H., Gispert, R., Lagache, G., et al. 2001, A\&A 372, 364

\bibitem[1998]{msx-irdc}
Egan, M. P., Shipman, R. F., Price, S. D., et al. 1998, ApJ 494, L199

\bibitem[1999]{ref-elbaz}
Elbaz, D., Cesarsky, C. J., Fadda, D., et al. 1999, A\&A 351, 37

\bibitem[1994]{DENIS}
Epchtein, N., de Batz, B., Copet, E., et al. 1994, Ap\&SS 217, 3

\bibitem[1997]{ref-DENIS2}
Epchtein, N., de Batz, B., Capoani, L., et al. 1997, The Messenger 87, 27

\bibitem[2000]{felli00}
Felli, M., Comoretto, G., Testi, L., Omont, A., \& Schuller, F. 2000,
A\&A 362, 199

\bibitem[2002]{felli02}
Felli, M., Testi, L., Schuller, F., \& Omont, A. 2002, A\&A 392, 971

\bibitem[2000]{Fouque}
Fouqu\'e, P., Chevallier, L., Cohen, M., et al. 2000, A\&AS 141, 313

\bibitem[1999]{glass99}
Glass, I. S., Ganesh, S., Alard, C., et al. 1999, MNRAS 308, 127

\bibitem[1998]{ref-gbpp}
Hammersley, P. L., Jourdain de Muizon, M., Kessler, M. F., et al.
1998, A\&AS 128, 207

\bibitem[2001]{hennebel}
Hennebelle, P., P\'erault, M., Teyssier, D., \& Ganesh, S. 2001, A\&A 365, 598

\bibitem[2003]{ref-jiang}
Jiang B. W., Omont A., Ganesh S., Simon G., Schuller F.,
A\&A in press, astro-ph/0302411

\bibitem[1996]{ISO_sat}
Kessler, M. F., Steinz, J. A., Anderegg, M. E., et al. 1996, A\&A 315, L27

\bibitem[2002]{msx-YSO}
Lumsden, S. L., Hoare, M. G., Oudmaijer, R. D., \& Richards, D. 2002,
MNRAS 336, 621

\bibitem[1994]{msx-overview}
Mill, J. D., O'Neil, R. R., Price, S., et al. 1994, Journal of Spacecraft
and Rockets 31, 900

\bibitem[2000]{ref-mamd}
Miville-Desch\^enes, M.-A., Boulanger, F., Abergel, A., \& Bernard, J.-P.
2000, A\&AS 146, 519

\bibitem[1998]{nordh}
Nordh, L., Olofsson, G., Bontemps, S., et al. 1998, in ASP Conf. Ser. 132,
Star Formation with the Infrared Space Observatory, ed. J. Yun \& R. Liseau, 
127

\bibitem[2003]{ref-ojha}
Ojha D. K., Omont A., Schuller F., et al. 2003, A\&A in press

\bibitem[1999]{omont99}
Omont, A., Ganesh, S., Alard, C., et al. 1999, A\&A 348, 755

\bibitem[2003]{Omont2002}
Omont, A., Gilmore, G. F., Alard, C., et al. 2003, A\&A, this issue

\bibitem[2002]{ref-ortiz}
Ortiz, R., Blommaert, J. A. D. L., Copet, E., et al. 2002, A\&A 388, 279

\bibitem[1997]{ref-CIA}
Ott, S., Abergel, A., Altieri, B., et al. 1997, Design and Implementation
of CIA, the ISOCAM Interactive Analysis System. In ASP Conf. Ser. 125, ed.
G. Hunt \& H. E. Payne, 34

\bibitem[2002]{Ott}
Ott, S. 2002, PhD Thesis, Paris VI - Pierre et Marie Curie University

\bibitem[1996]{perault}
P\'erault, M., Omont, A., Simon, G., et al. 1996, A\&A 315, L165

\bibitem[2001]{ref-MSX}
Price, S. D., Egan, M. P., Carey, S. J., Mizuno, D. R., \& Kuchar, T. A.
2001, AJ 121, 2819

\bibitem[1999]{ref-elais}
Rowan-Robinson, M., Oliver, S., Efstathiou, A., et al. 1999, in The Universe as
Seen by ISO, ed. P. Cox \& M. F. Kessler., ESA-SP 427, 1011

\bibitem[2002]{phd-schuller}
Schuller, F. 2002, PhD Thesis, Paris VI - Pierre et Marie Curie University

\bibitem[2000]{variable}
Schultheis, M., Ganesh, S., Glass, I. S., et al. 2000, A\&A 362, 215
  
\bibitem[1997]{ref-2mass}
Skrutskie, M. F., Schneider, S. E., Stiening, R., et al. 1997, in The Impact
of Large Scale Near-IR Sky Surveys, ed. F. Garzon et al. (Dordrecht: Kluwer), 25

\bibitem[1998]{ref-vision}
Starck, J.-L. 1998, in Les Houches Summer School on 'Infrared Astronomy from Space:
Today and Tomorrow'

\bibitem[1998]{ref-vision2}
Starck, J.-L., Murtagh, F., \& Bijaoui, A. 1998, {\em Image
Processing and Data Analysis: The Multiscale Approach},
Cambridge University Press

\bibitem[1997]{testi97}
Testi, L., Felli, M., Omont, A., et al. 1997, A\&A 318, L13 

\bibitem[2001]{zavagno}
Zavagno, A., \& Ducci, V. 2001, A\&A 371, 312

\end{thebibliography}
\end{document}